\documentclass[
nofootinbib,
noeprint,
amsmath,amssymb,
aps,
nobibnotes,
twocolumn,
floatfix,
]{revtex4-1}

\usepackage[utf8]{inputenc}
\usepackage{graphicx}
\usepackage{float}
\usepackage{epstopdf}
\usepackage{braket}

\usepackage{amsmath}
\usepackage{amssymb}
\usepackage{bm}
\usepackage{txfonts}

\usepackage{multirow}
\usepackage{mathptmx} 
\usepackage{siunitx}
\usepackage{caption}
\usepackage{subcaption}
\usepackage{xargs}                    
\usepackage[pdftex,dvipsnames]{xcolor}
\usepackage{datetime}
\ddmmyyyydate

\setcitestyle{super}

\usepackage[colorlinks=true, urlcolor=blue, linkcolor=blue, citecolor=blue]{hyperref}

\begin{document}
\renewcommand{\thefootnote}{\roman{footnote}}

\title{A quantum network of entangled optical atomic clocks}
\author{B. C. Nichol$^*$}\footnote[0]{*These authors contributed equally. \\Email:bethan.nichol@balliol.ox.ac.uk, \\{raghavendra.srinivas@physics.ox.ac.uk~}}
\author{R. Srinivas$^*$}
\author{D. P. Nadlinger}
\author{P. Drmota}
\author{D. Main}
\author{G. Araneda}
\author{C. J. Ballance}
\author{D. M. Lucas \\
\normalsize{Department of Physics, University of Oxford, Clarendon Laboratory, Parks Road, Oxford OX1 3PU, U.K.}\\
}
\date{\today}

\begin{abstract}
Optical atomic clocks are our most precise tools to measure time and frequency~\cite{Brewer2019, Oelker2019, Ludlow2015}. They enable precision frequency comparisons between atoms in separate locations to probe the space-time variation of fundamental constants~\cite{Rosenband2008}, the properties of dark matter~\cite{Derevianko2014, Safranova2018}, and for geodesy~\cite{Rosenband2008, Mehlstaubler2018, McGrew2018}. Measurements on independent systems are limited by the standard quantum limit (SQL); measurements on entangled systems, in contrast, can surpass the SQL to reach the ultimate precision allowed by quantum theory --- the so-called Heisenberg limit. While local entangling operations have been used to demonstrate this enhancement at microscopic distances~\cite{Meyer2001, Leibfried2004, Roos2006, Megidish2019, Manovitz2019, Pedrozo2020}, frequency comparisons between remote atomic clocks require the rapid generation of high-fidelity entanglement between separate systems that have no intrinsic interactions. We demonstrate the first quantum network of entangled optical clocks using two  $^{88}\text{Sr}^+$ ions separated by a macroscopic distance ($\approx$2\,m), that are entangled using a photonic link~\cite{Monroe2014, Stephenson2020}. We characterise the entanglement enhancement for frequency comparisons between the ions. We find that entanglement reduces the measurement uncertainty by a factor close to $\sqrt{2}$, as predicted for the Heisenberg limit, thus halving the number of measurements required to reach a given precision. Practically, today's optical clocks are typically limited by dephasing of the probe laser~\cite{Clements2020}; in this regime, we find that using entangled clocks confers an even greater benefit, yielding a factor 4 reduction in the number of measurements compared to conventional correlation spectroscopy techniques~\cite{Hume2016, Clements2020}.
As a proof of principle, we demonstrate this enhancement for measuring a frequency shift applied to one of the clocks. Our results show that quantum networks have now attained sufficient maturity for enhanced metrology. This two-node network could be extended to additional nodes~\cite{Komar2014}, to other species of trapped particles, or to larger entangled systems via local operations.
\end{abstract}

\maketitle

Nonclassical states enable measurements beyond the standard quantum limit (SQL)~\cite{Wineland1992, Wineland1994, Degen2017}. For example, quantum-enhanced measurements have been used for gravitational wave sensing~\cite{Caves1981, LIGO2019}, searches for dark matter~\cite{Malnou2019}, and force sensing~\cite{Gilmore2021}. While quantum networks have been used for quantum cryptography~\cite{Kimble2008, Gisin2002}, quantum computation~\cite{Monroe2013}, and verifications of quantum theory~\cite{Hensen2015}, they could potentially be used for enhanced metrology by distributing entanglement between remote systems. This enhancement is particularly important for optical atomic clock comparisons, where the number of measurements required to reach the noise floor is presently limited by the single-shot measurement uncertainty set by the SQL. Harnessing entanglement to move beyond the SQL and reach precision floors faster will enable the detection of phenomena on shorter timescales, and reveal previously undetectable signals by reducing the demands on the stability of the system.

\begin{figure*}[!t]
\centering
{
\includegraphics[width=2\columnwidth]{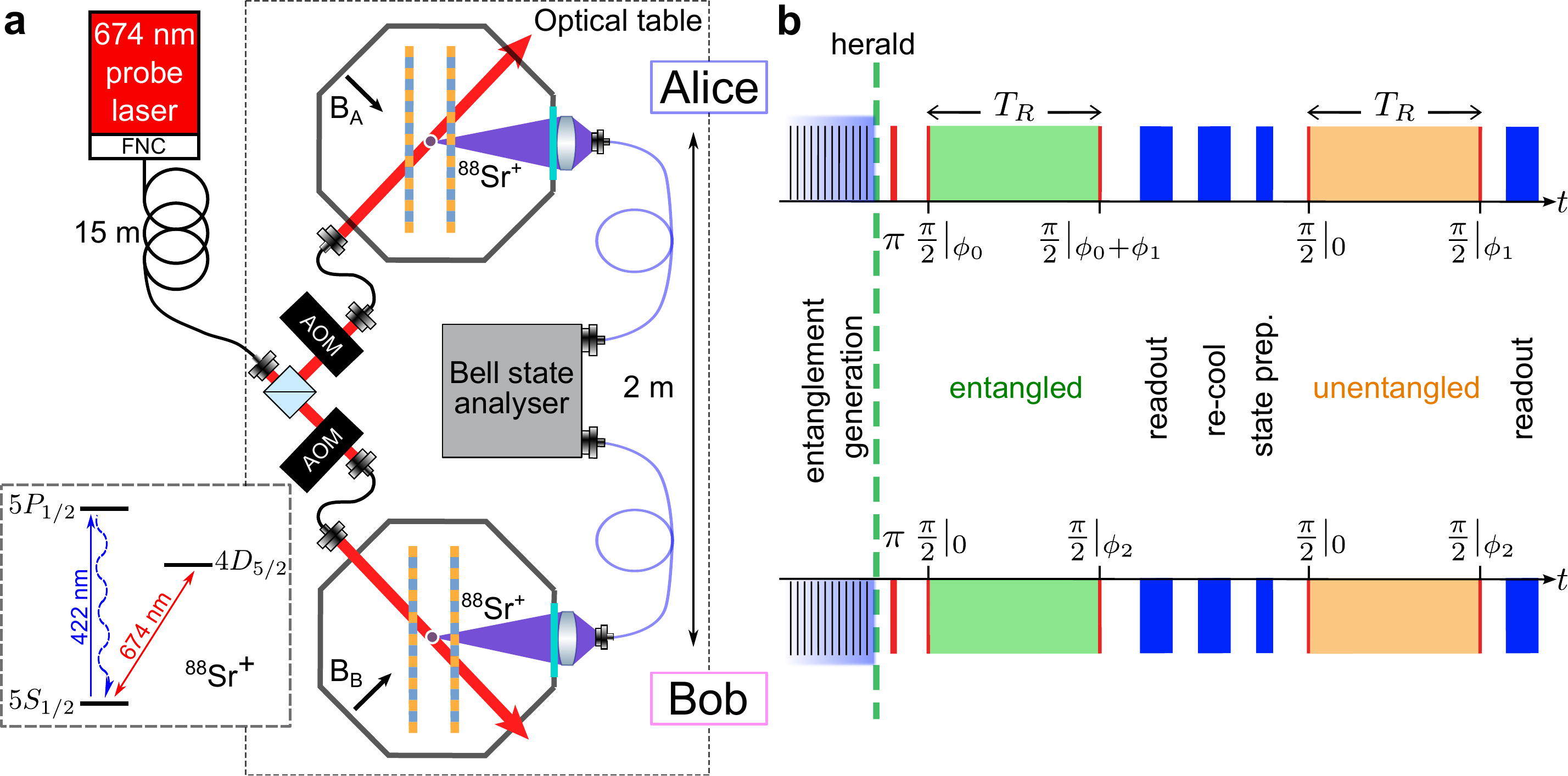}}
\caption{\label{fig_setup}\textbf{Network of entangled optical clocks.} \textbf{a}, Experimental apparatus. The network comprises two trapped-ion systems, Alice and Bob, separated by 2\,m, each containing a single $^{88}$Sr$^+$ ion (not to scale). We use a photonic link to generate remote entanglement; spontaneously emitted 422\,nm photons are transmitted via optical fibres to a Bell state analyser where a measurement projects the ions into an entangled state within the $S_{1/2}$ manifold. We map this entanglement to the $S_{1/2}\leftrightarrow D_{5/2}$ optical clock transition using a common 674\,nm laser, which has fibre-noise cancellation (FNC) from the laser system to the optical table. This laser is also used to probe the clock transition. Each trap has an independent acousto-optical modulator (AOM) for switching, frequency and phase control. A magnetic field of $\sim0.5\,$mT, indicated by $B_{A/B}$, provides a quantisation axis in each trap. The relevant energy levels of $^{88}$Sr$^+$ are shown in the inset. \textbf{b}, Experimental pulse sequence. In a single experimental sequence, we measure the ion states after Ramsey experiments on both the entangled and unentangled states. Entanglement generation, using simultaneous 422\,nm excitation pulses, is repeatedly attempted until a coincident two-photon detection at the Bell state analyser heralds entanglement. A $674$\,nm $\pi$-pulse then maps the entanglement to the optical clock transition. We simultaneously perform a Ramsey experiment on each ion with a total probe duration of $T_R$; spin-echo pulses during $T_R$ are not shown. The phase of the first $\pi/2$ pulse on Alice is $\phi_0=\phi+\pi$, where $\phi$ is the phase of the initial entangled state.  Following ion readout, cooling, and state preparation, this process is repeated for the unentangled state.}
\end{figure*}
 
The standard method for optical atomic clock comparisons requires the measurement of each atomic frequency relative to a laser. This laser is used to drive a narrow optical atomic transition and its relative frequency is determined by observing changes in the atomic state. This measurement is typically performed using a Ramsey experiment~\cite{Ramsey1950, Ramsey1957}, where a superposition of two states, $\ket{\downarrow}$ and $\ket{\uparrow}$, evolves for a duration $T_R$ in between two $\pi/2$ pulses. A difference frequency between the atom and the laser results in a relative phase between the two atomic states, which can be measured by repeated observations of the final state of the atom. For a single atom $i$, the expectation value of this measurement is 

\begin{align}
    \braket{\hat{\Pi}_i} = C_i\cos(\Delta_iT_{R} + \phi_i),
    \label{eq_single_atom}
\end{align}

\noindent where $\Delta_i=\omega_L-\omega_i$ is the detuning between the laser frequency, $\omega_L$, and the atomic resonance frequency, $\omega_i$. Here, $\phi_i$ is the phase of the second $\pi/2$ pulse with respect to the first pulse, $C_i$ is the signal contrast which is ideally 1 in the absence of decoherence, and ${\hat{\Pi}_i=\hat{\sigma}_{zi}=\ket{\uparrow}\bra{\uparrow}-\ket{\downarrow}\bra{\downarrow}}$ is the spin measurement operator. The corresponding frequency uncertainty for a single quantum measurement is

\begin{align}
    \label{eq_single_shot}
    \delta\Delta_i = \frac{\delta\braket{\hat{\Pi}_i}}{C_iT_R},
\end{align}

\noindent where $\delta\braket{\hat{\Pi}_i}$ is ideally limited by quantum projection noise~\cite{Itano1993}. Minimising this `single-shot' uncertainty minimises the number of measurements required to reach a given precision. To compare two clocks, we wish to measure the frequency difference, $\Delta_-=\Delta_1-\Delta_2=\omega_2-\omega_1$. For completely independent systems, two independent measurements of $\Delta_1$ and $\Delta_2$ are required. Assuming each measurement has the same uncertainty, the single-shot uncertainty of $\Delta_-$ is $\delta\Delta_{-,s}=\sqrt{2}\delta\Delta_{i}$. A direct measurement of $\Delta_-$ can be made by probing two atoms with simultaneous Ramsey experiments and measuring the correlated two-ion parity signal~\cite{Leibfried2004}
\begin{align}
    \braket{\hat{\Pi}} = P_+\cos(\Delta_+T_R+\phi_+) + P_-\cos(\Delta_-T_R+\phi_-),
    \label{eq_two_atom}
\end{align}
\noindent where $\Delta_\pm=\Delta_1 \pm \Delta_2$, $\phi_\pm=\phi_1 \pm \phi_2$, and $\hat{\Pi}=\hat{\sigma}_{z1}\hat{\sigma}_{z2}$. The probabilities of the two-atom state during the Ramsey delay being $\frac{1}{\sqrt{2}}(\ket{\downarrow\uparrow}+ e^{i\phi}\ket{\uparrow\downarrow})$ or $\frac{1}{\sqrt{2}}(\ket{\downarrow\downarrow}+ e^{i\phi'}\ket{\uparrow\uparrow})$ are given by $P_-$ and $P_+$ respectively, where $\phi$ and $\phi'$ are arbitrary phases. For two maximally-entangled atoms, we can set $P_-=1$ (and $P_+=0$) to measure $\Delta_-$ directly with maximum signal contrast, resulting in a single-shot uncertainty $\delta\Delta_{-, e}=\delta\Delta_i={\delta\Delta_{-,s}}/{\sqrt{2}}$. This uncertainty follows the expected Heisenberg scaling and is $\sqrt{2}$ lower compared with making two independent measurements~\cite{Giovannetti2006}.

With present optical clock technology, dephasing of the probe laser typically limits the uncertainty that can be achieved~\cite{Riis2004, Leroux2017}. For the single-atom measurements, laser phase noise effectively randomises $\phi_i$ in Eq.~\ref{eq_single_atom}, reducing the contrast $C_i$ and setting a practical limit on the duration $T_R$. In this regime, entanglement offers an additional advantage as $\phi_-$ has complete cancellation of common phase fluctuations and is only affected by differential phase noise between the two systems. If entanglement is not available as a resource, this insensitivity can also be accessed by using conventional correlation spectroscopy~\cite{Bize2000, Chwalla2007, Marti2018, Young2020}, which involves simultaneous measurements with a common probe laser and an unentangled two-atom state~\cite{Hume2016}, as recently demonstrated by Clements \textit{et al.}~\cite{Clements2020} for two macroscopically-separated clocks. However, in this scenario $P_+=P_-=\frac{1}{2}$ which sets the limit on the measurement uncertainty to $\delta\Delta_{-,u}=2\delta\Delta_i=\sqrt{2}\delta\Delta_{-,s}$, a factor of $\sqrt{2}$ worse than independent single-ion measurements. Entanglement enables one to combine this noise-insensitivity with the maximum measurement precision allowed by quantum mechanics.

\begin{figure*}[!t]
{\includegraphics[width=2\columnwidth]{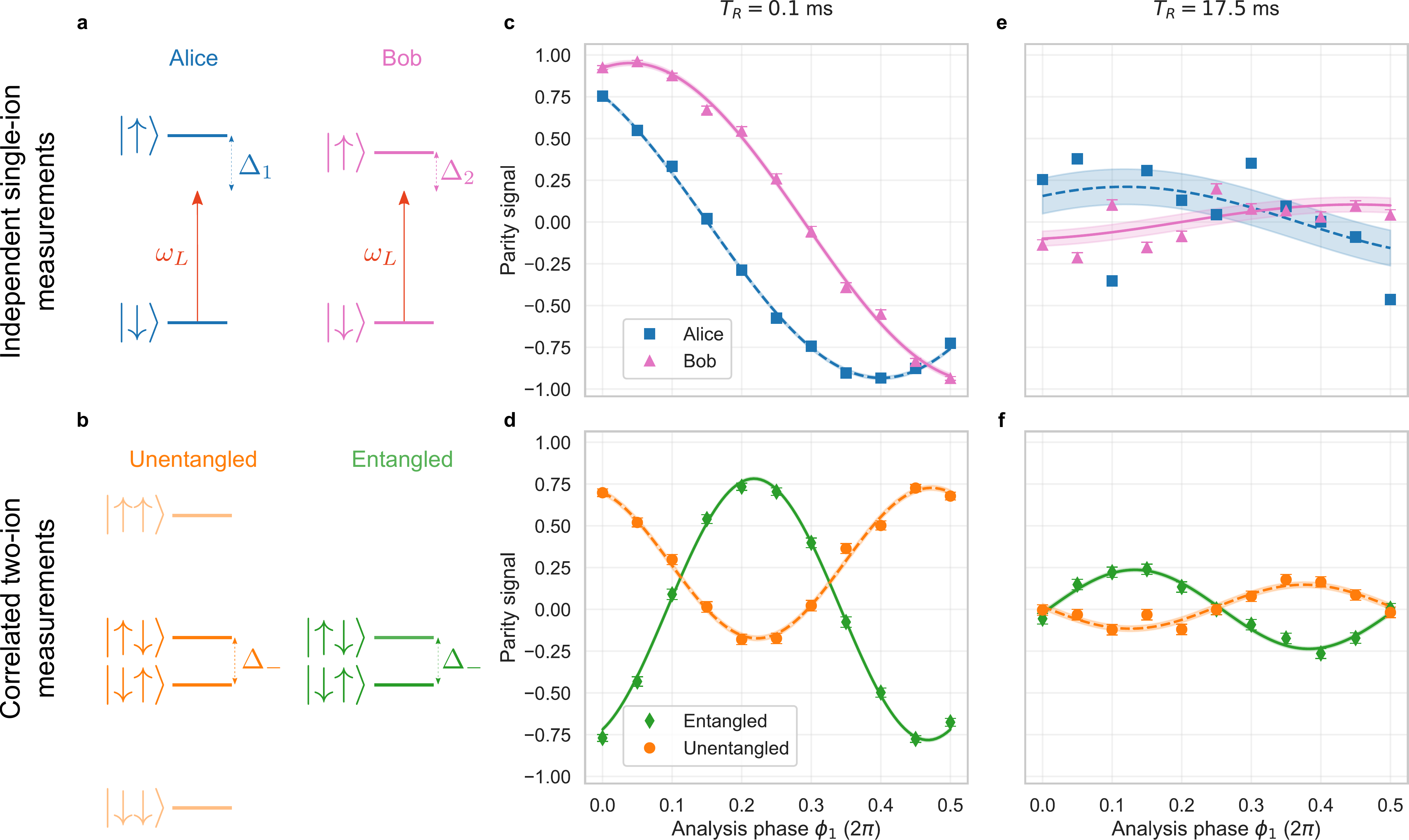}}
\centering
\caption{\label{fig_parity}\textbf{Spectroscopy with and without entanglement.} \textbf{a}, We perform Ramsey experiments on Alice's (blue) and Bob's (pink) atoms, using a probe laser with frequency $\omega_L$. These experiments can measure the atoms' detunings $\Delta_1$ and $\Delta_2$ between $\omega_L$ and the $\ket{\downarrow}\leftrightarrow\ket{\uparrow}$ transitions. Thus, two independent measurements are required to obtain the difference frequency $\Delta_-=\Delta_1-\Delta_2$. \textbf{b}, With entanglement, only a single measurement is required for $\Delta_-$ using the states $\ket{\downarrow\uparrow}$ and $\ket{\uparrow\downarrow}$. A single measurement can also be performed using unentangled states, at the cost of having population in the additional states $\ket{\downarrow\downarrow}$ and $\ket{\uparrow\uparrow}$ that does not contribute to the signal. For \textbf{c}-\textbf{f}, we scan the analysis phase $\phi_1=-\phi_2$ from 0 to $\pi$. We plot the single-ion \textbf{(c)} and two-ion parity \textbf{(d)} signals at a Ramsey duration of 0.1\,ms. Imperfect entangled-state generation and an increased effect from the imperfect spin-echo pulses reduce the contrast of the entangled state (green diamonds), compared to the single ion scans (blue squares and pink triangles). The two-ion signal from the unentangled state (orange circles) has about half the contrast of the entangled state signal; the y-offset from 0 arises from the term containing $\Delta_+$ in Eq.~\ref{eq_two_atom}. Similarly, we plot the single-ion \textbf{(e)} and two-ion parity \textbf{(f)} signals at a Ramsey duration of 17.5\,ms, by which point the $\Delta_+$ term has averaged to 0. At this duration, both the single-ion and two-ion signals are reduced due to qubit decoherence from differential magnetic field fluctuations, but the entanglement enhancement is still evident. The single-ion signals have almost no visibility and increased uncertainty in the fits due to sensitivity to common-mode laser phase noise. Experimental data are shown as points, with error bars calculated from projection noise. Fits to the data (lines) are shown with 68\% confidence intervals (shaded region), following Eq.~\ref{eq_single_atom} and Eq.~\ref{eq_two_atom} for the single-ion and two-ion signals respectively.}
\end{figure*}

The challenge in realising this entanglement enhancement for remotely-located atoms is that the entangled state needs to be generated with both high fidelity (to achieve $P_-\approx 1$) and high speed (to maximise the measurement duty cycle). This has prevented previous experimental demonstrations. Using our two-node trapped-ion quantum network~\cite{Stephenson2020}, we can create entangled states of two remote $^{88}$Sr$^+$ ions with a fidelity of 0.960(2) in an average duration of 8\,ms~\cite{Nadlinger2021}, sufficient to realise the first elementary network of entangled optical clocks. We compare the $^{88}$Sr$^+$ optical clock transition frequency using (i) independent measurements on each atom, and correlated measurements of both (ii) unentangled atoms, and (iii) entangled atoms. We characterise the enhancement gained from entanglement and, as a proof-of-principle, make an entanglement-enhanced measurement of a frequency shift applied to one of the atoms.

For these experiments, we drive the $5S_{1/2}\leftrightarrow 4D_{5/2}$ optical clock transition in each trapped-ion system (labelled Alice and Bob)  with light from a common 674\,nm laser as shown in Fig.~\ref{fig_setup}a. Entanglement generation, Doppler cooling, state preparation, and readout are performed using the 422\,nm $S_{1/2}\rightarrow P_{1/2}$ transition. The initial atom-atom entanglement is generated using qubit states within the $S_{1/2}$ manifold that are separated at radiofrequency by 14\,MHz~\cite{Stephenson2020}. To map this entanglement to the 445\,THz optical qubit in each atom, we use a resonant 674\,nm $\pi$-pulse on the $S_{1/2}\leftrightarrow D_{5/2}$ transition (see Supplementary Information {\S}\ref{sec_mapping}). We create the entangled state $\ket{\Psi}=\frac{1}{\sqrt{2}}(\ket{\downarrow\uparrow}\pm e^{i\phi}\ket{\uparrow\downarrow})$, where $\phi$ is set by the difference in optical paths and magnetic field strengths between the two systems, and $\pm$ depends on the measurement outcome of the Bell state analyser. The qubit states are $\ket{\downarrow}\equiv\ket{S_{1/2}, m_j={-1/2}}$, and $\ket{\uparrow}\equiv\ket{D_{5/2}, m_j={-3/2}}$, where $m_j$ is the projection of the angular momentum along the quantisation axis defined by a static magnetic field $\approx0.5$\,mT. We calibrate the phase of the initial $\pi/2$ pulse for Alice's atom, relative to $\phi$, to remain in the optimal entangled state for measuring $\Delta_-$ (see Supplementary Information {\S}\ref{sec_bellcal}). The data required for comparison of the three measurement methods, at a given Ramsey duration, is taken in a single experiment sequence as shown in Fig.~\ref{fig_setup}b.  We use the data from the unentangled state to obtain both the single-ion and two-ion parity signals. The dominant source of decoherence in this experiment is fluctuation of the qubit transition frequency due to magnetic field noise. To suppress this decoherence and enable longer Ramsey durations, we use active magnetic field stabilisation and a modified spin-echo sequence (see Supplementary Information {\S}\ref{sec_echo}).

We perform Ramsey experiments with durations from 0.1\,ms to 20\,ms for both the entangled and unentangled states. Figure~\ref{fig_parity} shows a comparison of these experiments at 0.1\,ms and 17.5\,ms. To measure the oscillation only from the term in Eq.~\ref{eq_two_atom} containing $\Delta_-$, we set $\phi_2=-\phi_1$ and scan $\phi_1$. At $T_R=0.1$\,ms, we observe a slight reduction in contrast for the unentangled ions, mainly due to imperfect spin-echo pulses; imperfect entanglement fidelity further reduces the contrast for the entangled state. At longer durations, qubit decoherence due to magnetic field noise reduces the contrast of all the parity signals (see Fig.~\ref{fig_parity_time}). The sensitivity of the single-ion signal to laser phase noise is evident from the additional reduction in the contrast. When the laser phase noise at the fibre output was intentionally increased by turning off the FNC, we observed high contrast for the two-ion signals at probe durations where the single-ion signals had decohered completely (see Fig.~\ref{fig_fnc_off}). This demonstrates a decoherence-free subspace encoded in two qubits separated by a macroscopic distance. In principle, our network could also be used to enhance measurements of $\Delta_+$, which is required to stabilise the laser frequency to the mean atomic frequency. However, the entangled state needed for this measurement has an increased sensitivity to laser dephasing (see Fig.~\ref{fig_even}), hence accessing this enhancement will require improvements in laser technology.

\begin{figure}[!h]
\includegraphics[width=1\columnwidth]{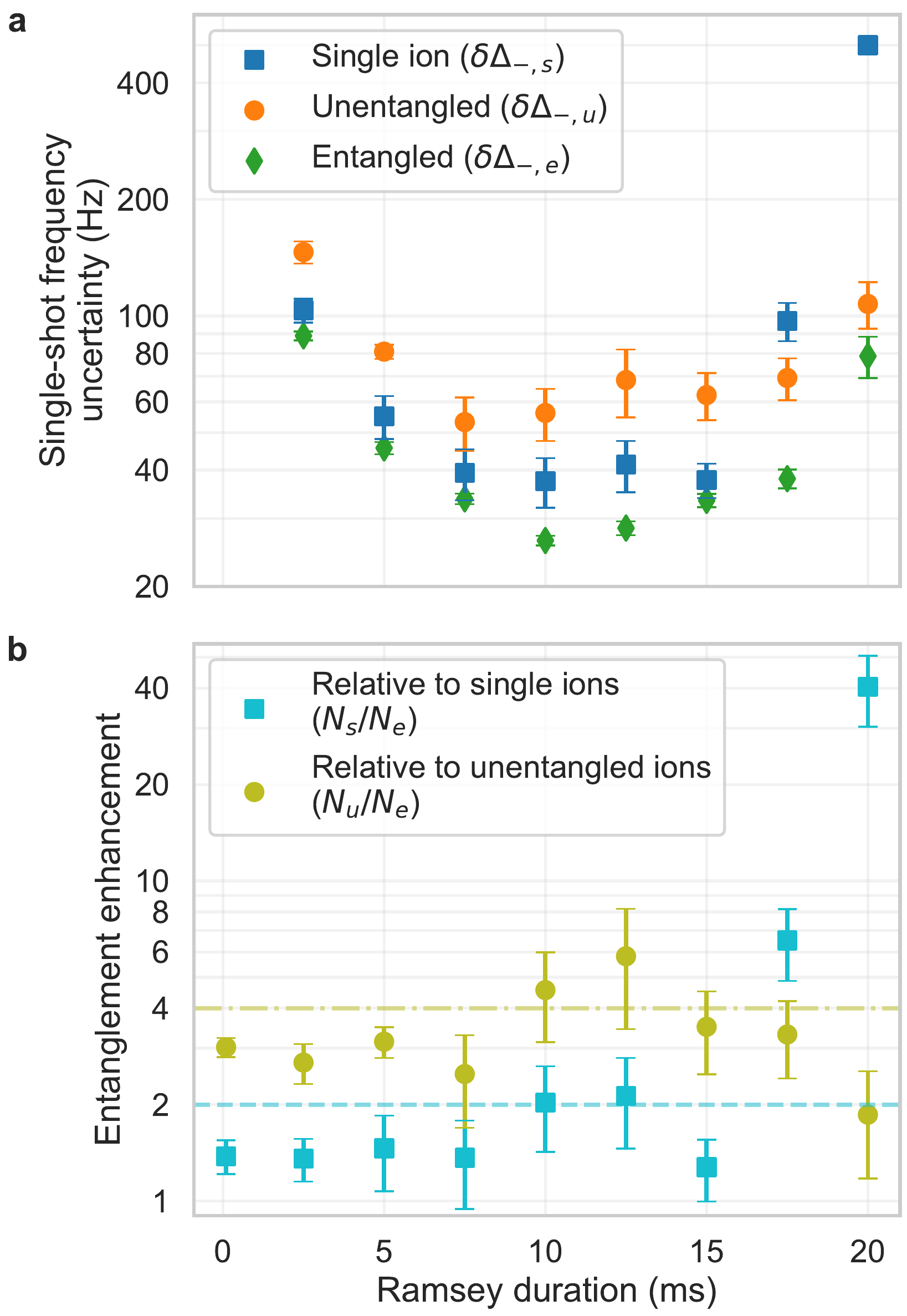}
\centering
\caption{\label{fig_enhancement} \textbf{Characterisation of entanglement enhancement.} \textbf{a}, Single-shot frequency uncertainties for single-ion ($\delta\Delta_{-,s}$), correlated unentangled ($\delta\Delta_{-,u}$), and entangled ($\delta\Delta_{-,e}$) measurements, versus Ramsey duration ($T_R$). The data at $T_R=0.1$\,ms are omitted for clarity. At all durations, the entangled state achieves the lowest single-shot uncertainty. At longer durations, the single-ion measurements have the highest uncertainty due to their sensitivity to laser dephasing. \textbf{b}, Entanglement enhancement versus Ramsey duration ($T_R$) relative to single-ion measurements (turquoise squares) and relative to measurements with two unentangled ions (olive circles). The theoretical enhancements are 2 and 4 for the single-ion (blue dashed line) and unentangled state (orange dash-dotted line), respectively. All error bars indicate 68\% confidence intervals.
}
\end{figure}

To characterise the enhancement from the entangled state for frequency comparisons, we plot the single-shot uncertainty for the single-ion and two-ion measurements versus the Ramsey duration (see Fig.~\ref{fig_enhancement}a). The uncertainty from the entangled state is lowest at all durations, with a minimum at 10\,ms. At longer durations, the reduction in contrast increases the net measurement uncertainty. We observe an advantage of probing two ions simultaneously at longer Ramsey durations, where the uncertainty for the single-ion measurements increases due to laser dephasing. We define the entanglement enhancement as the ratio of the number of measurements $N_{s/u}$ required to reach a given precision without entanglement, to the number of measurements $N_e$ required using the entangled state. The achievable precision using a particular method is 

\begin{align}
\sigma_{s/u/e}=\frac{\delta\Delta_{-, s/u/e}}{\sqrt{N_{s/u/e}}},    
\end{align}
\noindent where $s$, $u$, and $e$ correspond to measurements with single ions, unentangled ions, or entangled ions, respectively. Thus, the entanglement enhancement is equivalent to ${N_{s/u}/N_e=(\delta\Delta_{-,s/u}/\delta\Delta_{-,e})^2}$. In Fig.~\ref{fig_enhancement}b we plot the entanglement enhancement versus the Ramsey duration. Relative to the single-ion measurements, we observe an enhancement which is initially close to the expected factor of $2$, but increases at longer durations. In this regime, correlated measurements using two unentangled ions have a reduced uncertainty compared to the the single-ion measurements. Entanglement yields an even greater enhancement: we observe close to a factor of $4$ enhancement relative to the unentangled state at all durations.

\begin{figure}[!t]
\includegraphics[width=1\columnwidth]{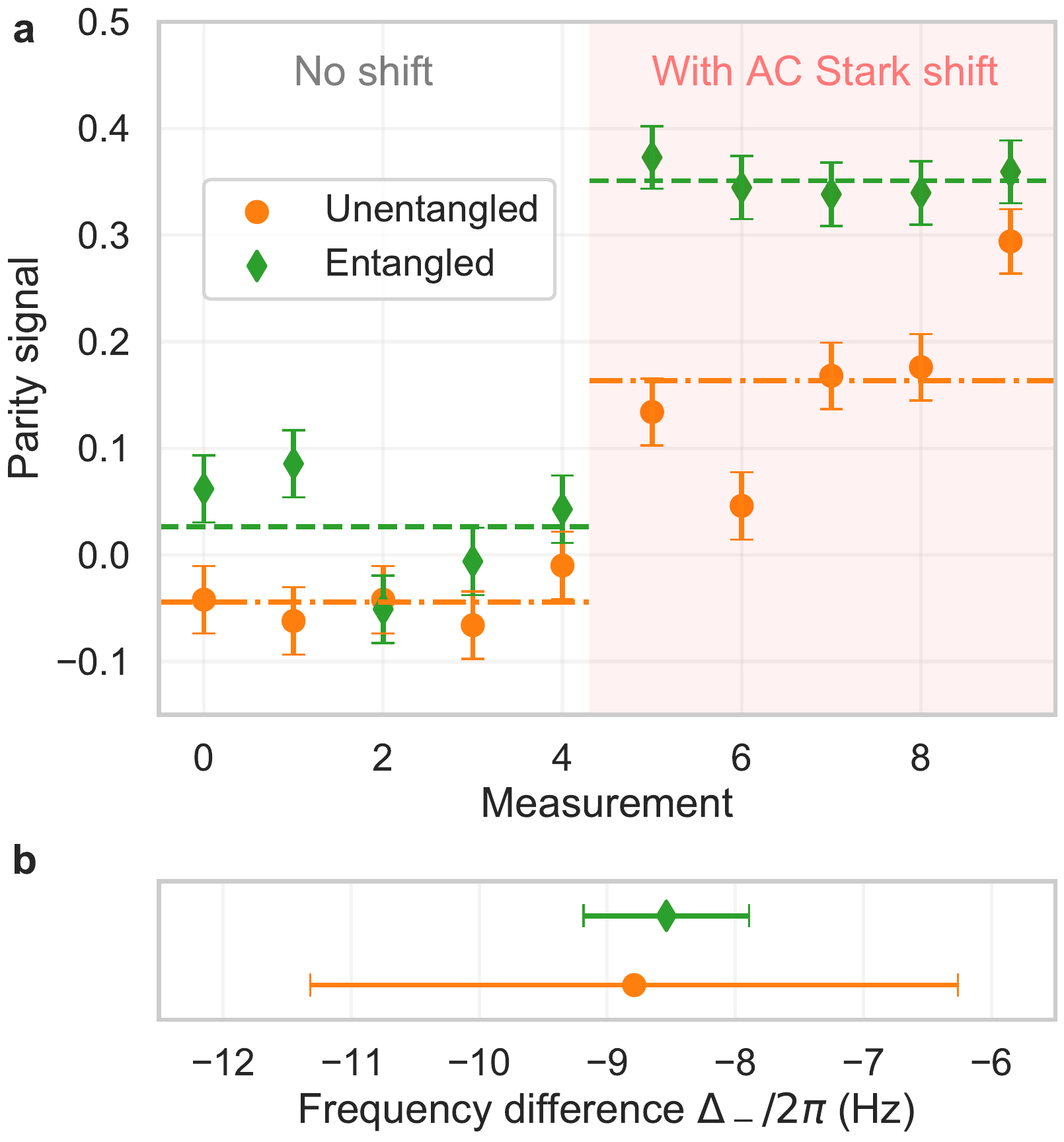}
\centering
\caption{\label{fig_shift} \textbf{Detection of clock-clock frequency difference with and without entanglement.} We plot the two-ion parity signal with (green diamonds) and without (orange circles) entanglement at a Ramsey duration of 15\,ms, choosing the analysis phase $\phi_1$ to sit at the steepest slope of the parity signal (see Fig.~\ref{fig_stark_cal}). The average parity signal with and without the shift are shown for the entangled (green dashed line) and unentangled state (orange dash-dotted line). The first five points are without any frequency shift, the next five points (shaded area) are with a shift applied to Alice's ion. The change in parity signal for the entangled state is about twice as large as for the unentangled states. Error bars indicate 68\% confidence intervals, given by quantum projection noise. \textbf{b}, Measured frequency difference with and without entanglement. From the change in the parity signal measured in \textbf{a}, we determine frequency differences of $-8.5\pm0.6$\,Hz and $-8.8\pm2.5$\,Hz, with and without entanglement, respectively. Error bars indicate 68\% confidence intervals.}
\end{figure}

Finally, as a proof of principle, we use entanglement to enhance the measurement of an applied frequency difference between the two ions. This frequency difference arises from an AC Stark shift due to a far-detuned $674$\,nm beam which illuminates Alice's ion throughout the 15\,ms Ramsey delay. As shown in Fig.~\ref{fig_shift}a, we see about a factor 2 increase in the two-ion parity signal, compared with the unentangled state, when the shift is applied. From the parity response, we measure a frequency difference $\Delta_-$ of $-8.8\pm2.5$\,Hz with the unentangled state, and $-8.5\pm0.6$\,Hz with the entangled state (see Fig.~\ref{fig_shift}b). The measurement uncertainty without entanglement is higher than the expected factor of 2 compared to the measurement with entanglement. This is likely due to an increased uncertainty in the parity signal for the unentangled state due to the term corresponding to $P_+$ in Eq.~\ref{eq_two_atom}, which has not yet completely averaged to zero at this Ramsey duration. This results in an above-statistical scatter in the unentangled measurements, as can be seen in Fig.~\ref{fig_shift}a.

In conclusion, we have demonstrated enhanced frequency comparisons using a quantum network of two entangled trapped-ion atomic clocks. The high fidelity and speed of entanglement generation in our network, which give a large signal and an efficient duty cycle, show that entangled clocks can already offer a practical enhancement for metrology. Compared to probe durations of $\sim$500\,ms used in state-of-the-art optical clocks~\cite{Oelker2019}, our entanglement generation duration of 9\,ms is negligible. We were restricted to the use of short probe durations (compared to the limit set by the $^{88}$Sr$^+$ $4D_{5/2}$ lifetime~\cite{Sahoo2006} of $\approx400\,$ms) by qubit decoherence due to magnetic field fluctuations. This decoherence limited our absolute measurement precision to a fractional frequency uncertainty of $\sim 10^{-15}$, well short of the state of the art for optical clocks~\cite{Brewer2019}. The magnetic field fluctuations could be reduced significantly through the use of superconducting solenoids~\cite{Gabrielse1988}, mu-metal shielding~\cite{Ruster2016}, or more advanced dynamical decoupling schemes~\cite{Aharon2019}. While our demonstration used single $^{88}$Sr$^+$ ions, whose simple level structure enables fast entanglement generation, the remote entanglement could in principle be mapped to any ion species via quantum logic operations~\cite{Schmidt2005}, with negligible loss of fidelity or speed~\cite{Hughes2020}. For example, we could choose an ion with a transition that has a reduced magnetic field sensitivity~\cite{Brewer2019}, a narrower linewidth, or an increased sensitivity to fundamental constants~\cite{Safranova2018}. Increasing the distance between nodes is important for remote sensing and geodesy; longer fibres could be used at the cost of a reduced entanglement rate due to fibre losses at 422~nm, or by downconversion to telecom wavelengths~\cite{Wright2018}. A larger network of atomic clocks can reduce the measurement uncertainty further~\cite{Komar2014}; we could increase the the number of ions in the network either by using local entangling operations between additional ions at each node, or by increasing the number of nodes using additional photonic links. Our demonstration provides the first building block towards such a network that could operate beyond the SQL, at the fundamental Heisenberg limit. \\

\noindent {\bf Acknowledgements:} We thank E.R. Clements, R.M. Godun, D.B. Hume, and A.M. Steane for helpful discussions and insightful comments on the manuscript. We would like to thank Sandia National Laboratories for supplying the HOA2 ion traps used in these experiments.
\noindent {\bf Funding:} This work was supported by the UK EPSRC Hub in Quantum Computing and Simulation (EP/T001062/1), the E.U. Quantum Technology Flagship Project AQTION (No.  820495), and C.J.B.’s UKRI Fellowship (MR/S03238X/1). B.C.N. acknowledges funding from the UK National Physical Laboratory.
\noindent{\bf Author contributions:} D.P.N., B.C.N., P.D., D.M., G.A., and R.S. built and maintained the experimental apparatus; R.S. conceived the experiments; B.C.N. and R.S. carried out the experiments, assisted by D.P.N., P.D., D.M., and G.A.; B.C.N., R.S., and D.M.L. analyzed the data; B.C.N. and R.S. wrote the manuscript with input from all authors; C.J.B. and D.M.L. secured funding for the work and supervised the work.
\noindent{\bf Competing interests:} C.J.B. is a director of Oxford Ionics. The  remaining authors declare no competing interests.
\noindent{\bf Data availability:} Source data for all plots are available. All other data are available from the corresponding authors upon reasonable request.

\clearpage
\section*{Supplementary Information}

\setcounter{equation}{0}
\setcounter{figure}{0}
\renewcommand{\theequation}{S\arabic{equation}}
\renewcommand{\thefigure}{S\arabic{figure}}

\subsection{Mapping remote entanglement to optical qubit}
\label{sec_mapping}

The initial entanglement is generated using the $\ket{S_{1/2}, m_J=\pm1/2}$ states that are separated by 14\,MHz~\cite{Stephenson2020}. To map this entanglement to the optical qubit on each atom, we use a $\pi$-pulse on the $\ket{S_{1/2}, m_J=+1/2}\leftrightarrow\ket{D_{5/2}, m_J=-3/2}$ transition. This sequence creates the entangled Bell state

\begin{align}
    \ket{\Psi}=\frac{1}{\sqrt{2}}(\ket{\downarrow\uparrow}\pm e^{i\phi}\ket{\uparrow\downarrow}),
\end{align}

\noindent where $\ket{\downarrow}\equiv\ket{S_{1/2}, m_j={-1/2}}$, and $\ket{\uparrow}\equiv\ket{D_{5/2}, m_j={-3/2}}$.

\subsection{Bell state calibration}
\label{sec_bellcal}

We measure the frequency difference between the atoms $\Delta_-=\Delta_1-\Delta_2=\omega_2-\omega_1$, where $\Delta_i=\omega_L-\omega_i$. The laser and atomic frequencies are denoted by $\omega_L$ and $\omega_i$ respectively. To measure $\Delta_-$ using entangled atoms, we require the state ${\ket{\Psi}=\frac{1}{\sqrt{2}}(\ket{\downarrow\uparrow}\pm e^{i\phi}\ket{\uparrow\downarrow})}$. Entanglement can also improve measurements of $\Delta_+=\Delta_1+\Delta_2$, which can be used to stabilise the laser frequency to the mean atomic frequency. This measurement requires the entangled state ${\ket{\Phi}=\frac{1}{\sqrt{2}}(\ket{\downarrow\downarrow}\pm e^{i\phi'}\ket{\uparrow\uparrow})}$.

\begin{figure}[!h]
\includegraphics[width=1\columnwidth]{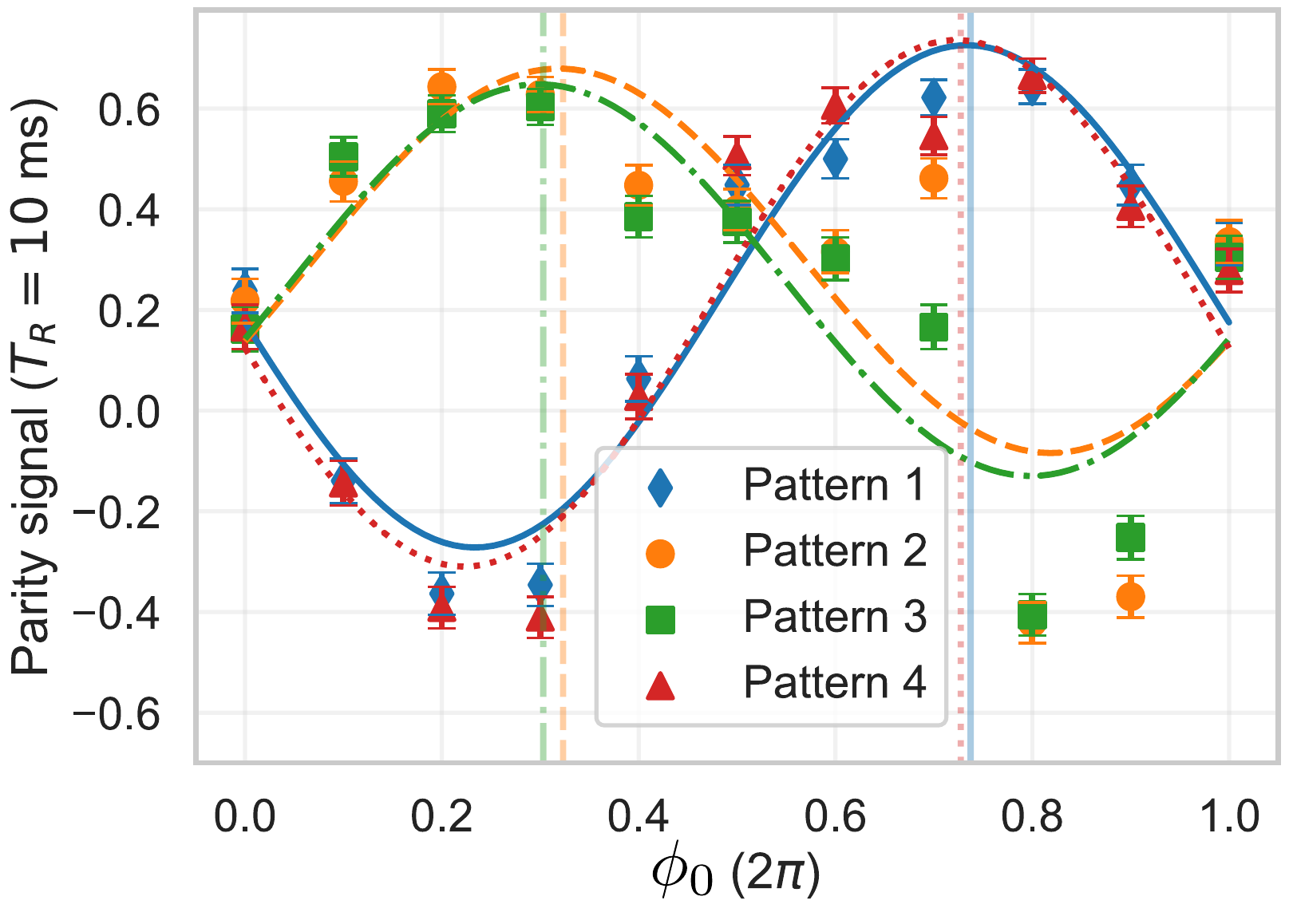}
\centering
\caption{\label{fig_bellstate} Calibration of $\phi_0$, the relative phase of the Ramsey experiments in Alice with respect to Bob. For a Ramsey duration of 10\,ms, and fixed analysis phases $\phi_1$ and $\phi_2$, we plot the two-ion parity versus $\phi_0$ for each of the 4 detector patterns in the Bell state analyser~\cite{Stephenson2020}. The optimal $\phi_0$ corresponds to the peak of parity oscillation (vertical lines). Patterns 1 (blue diamonds) and 4 (red triangles) are approximately $\pi$ out of phase from patterns 2 (orange circles) and 3 (green squares), corresponding to the phase offset of the resulting entangled state.}
\end{figure}

\begin{figure}[!t]
\includegraphics[width=1\columnwidth]{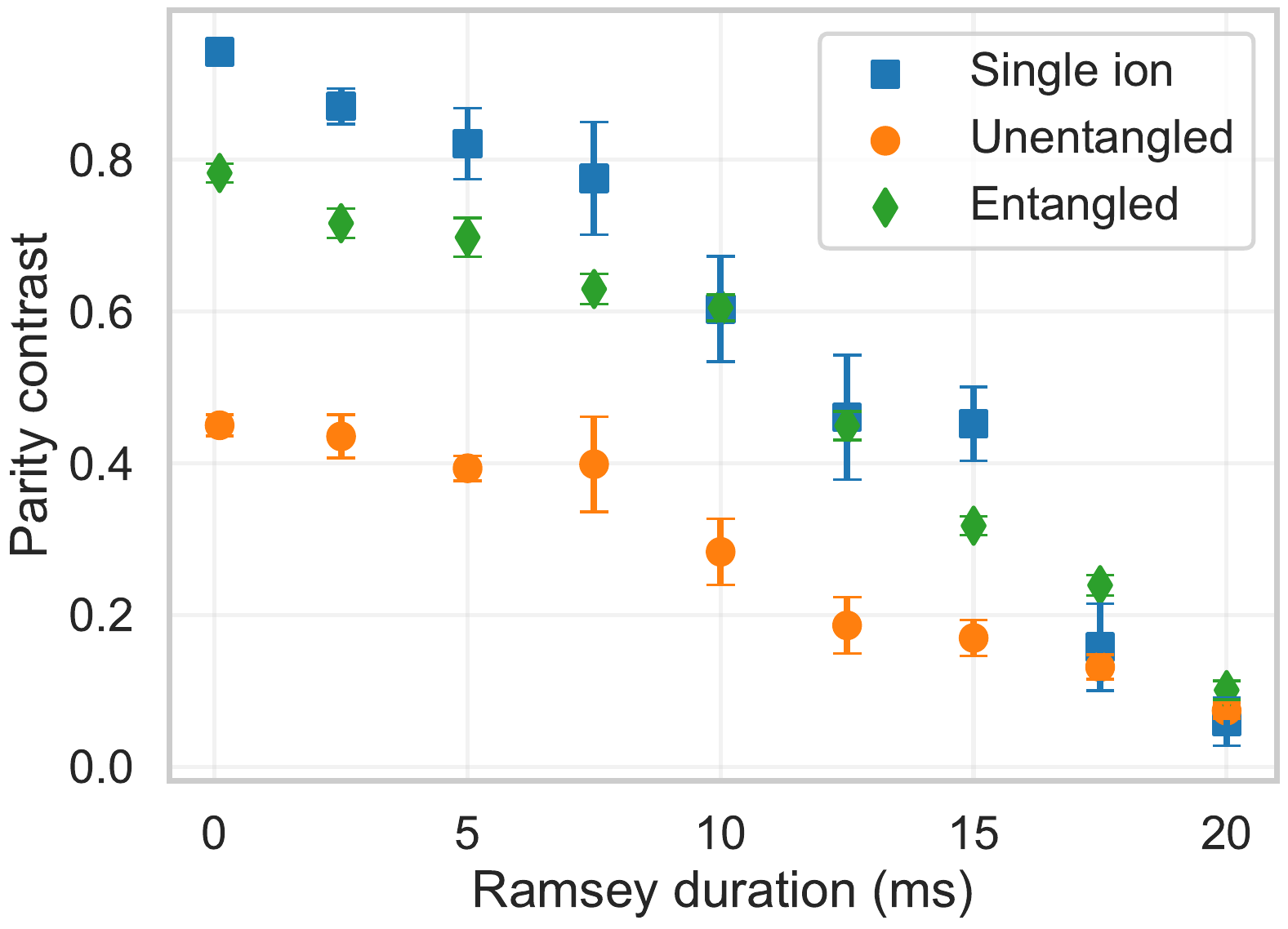}
\centering
\caption{\label{fig_parity_time}  Parity contrast versus Ramsey duration. For each Ramsey duration, we perform phase scans as shown in Fig.~\ref{fig_parity} and measure the contrast. The single-ion measurements are the average contrast from Alice and Bob. For single-ion (blue squares), unentangled (orange circles), and entangled (green diamonds) measurements, the contrast decays due to qubit decoherence. The contrast for the single-ion measurements also decays due to laser dephasing.}
\end{figure}

\begin{figure}[!h]
\includegraphics[width=1\columnwidth]{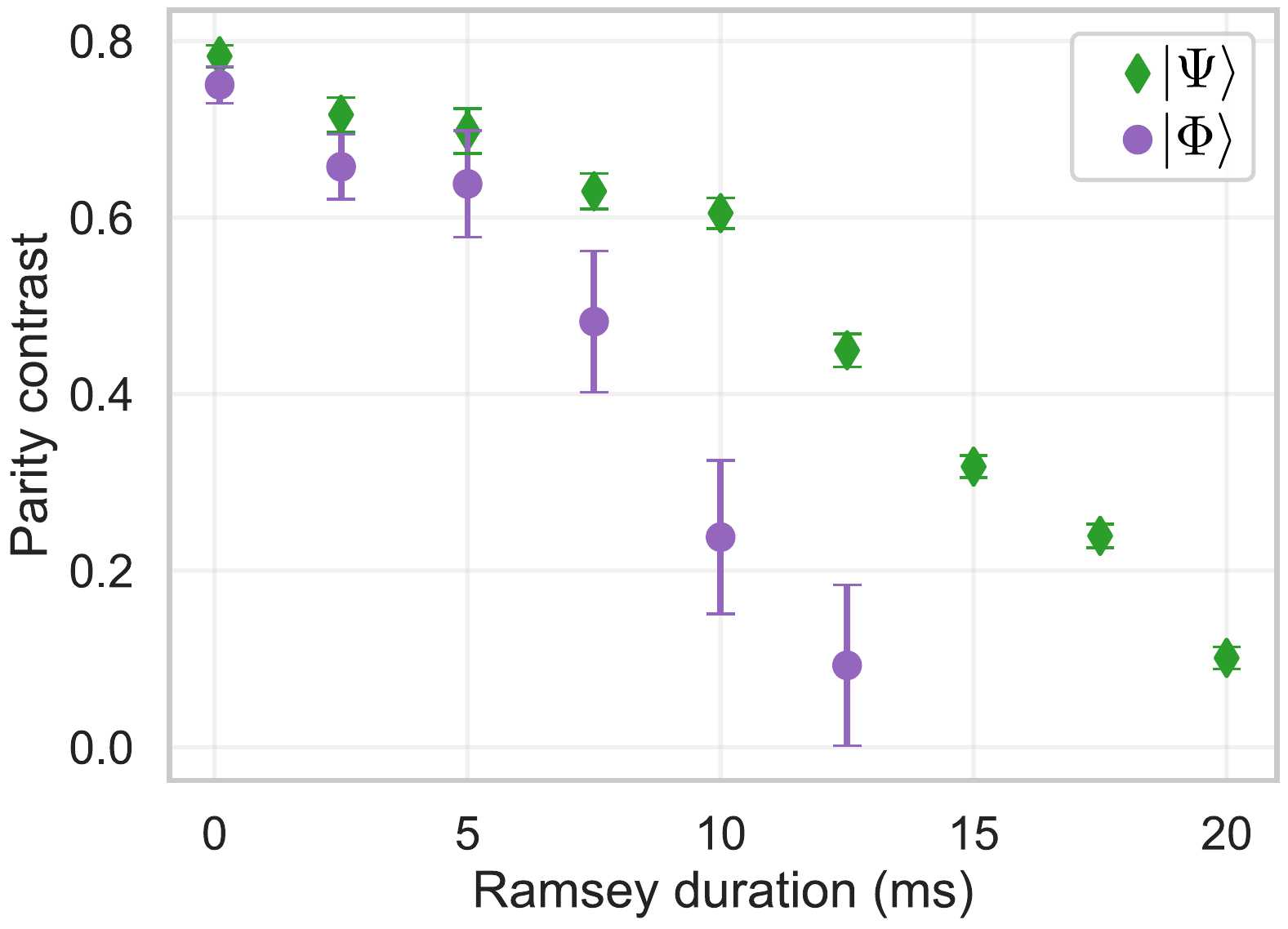}
\centering
\caption{\label{fig_even} Two-ion parity contrast versus Ramsey duration for $\ket{\Psi}=\frac{1}{\sqrt{2}}(\ket{\downarrow\uparrow}\pm e^{i\phi}\ket{\uparrow\downarrow})$ and ${\ket{\Phi}=\frac{1}{\sqrt{2}}(\ket{\downarrow\downarrow}\pm e^{i\phi'}\ket{\uparrow\uparrow})}$. Compared to $\ket{\Psi}$ (green diamonds), $\ket{\Phi}$ (purple circles) has a faster decay in contrast due to an increased sensitivity to laser dephasing. All error bars indicate 68\% confidence intervals.}
\end{figure}

We initially create the entangled state ${\ket{\Psi}=\frac{1}{\sqrt{2}}(\ket{\downarrow\uparrow}\pm e^{i\phi}\ket{\uparrow\downarrow})}$, where $\phi$ is set by optical paths in the apparatus, and $\pm$ corresponds to the specific detector pattern in the Bell state analyser that heralds entanglement~\cite{Stephenson2020}. To remain in the optimal entangled state for measuring $\Delta_-$, we adjust the phase of the Ramsey experiments ($\phi_0$ in Fig~\ref{fig_setup}b) in Alice with respect to Bob to be $\phi_0=\phi+\pi$. Sample calibration data is shown in Fig.~\ref{fig_bellstate}, where we determine the optimal $\phi_0$ required for each of the 4 possible detector patterns. This phase is updated in real time within the experimental sequence to acquire data from all heralds. Figure~\ref{fig_parity_time} shows the parity contrasts versus time for the single-ion measurements, the two-ion unentangled state and the two-ion entangled state $\ket{\Psi}$. By setting $\phi_0=\phi$, we can transform $\ket{\Psi}$ to $\ket{\Phi}$. We performed Ramsey experiments using $\ket{\Phi}$ as shown in Fig.~\ref{fig_even}. This state has a faster decay in contrast due to increased sensitivity to laser dephasing.

\subsection{Echo sequence}
\label{sec_echo}

\begin{figure}[!h]
\includegraphics[width=0.8\columnwidth]{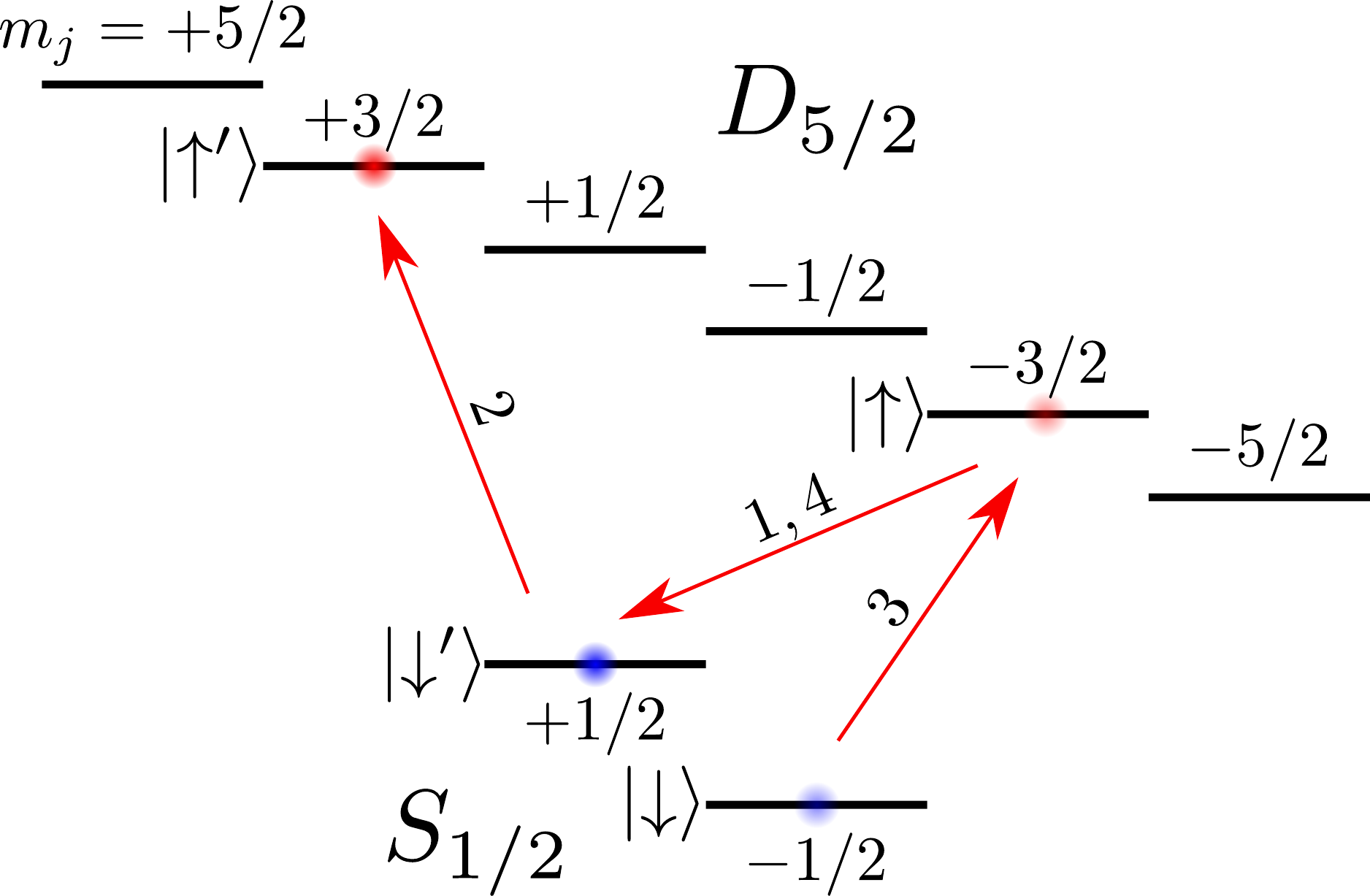}
\centering
\caption{\label{fig_levels} Modified spin-echo sequence. We map the $\ket{\downarrow}$ and $\ket{\uparrow}$ qubit states to the $\ket{\downarrow'}$ and $\ket{\uparrow'}$ states following the pulse sequence described in the text. The primed states have equal and opposite magnetic field sensitivities to the unprimed states.}
\end{figure}

We alternate between the transitions ${\ket{\downarrow}\equiv\ket{S_{1/2}, m_j={-1/2}}\leftrightarrow\ket{\uparrow}\equiv\ket{D_{5/2}, m_j={-3/2}}}$ and $\ket{\downarrow'}\equiv{\ket{S_{1/2}, m_j={+1/2}}\leftrightarrow\ket{\uparrow'}\equiv\ket{D_{5/2}, m_j={+3/2}}}$, which have magnetic field sensitivities of $+$11.2\,MHz/mT and $-$11.2\,MHz/mT, respectively. Thus, by alternating between these transitions, we can `echo out' slow qubit frequency drifts due to fluctuations in the magnetic field, while still measuring a shift of the centre-of-gravity of the 674\,nm transition. To map $\ket{\uparrow}\rightarrow\ket{\uparrow'}$ and $\ket{\downarrow}\rightarrow\ket{\downarrow'}$, we perform the following echo sequence using 674\,nm $\pi$-pulses resonant with the required transition, as illustrated in Fig.~\ref{fig_levels}: 

\begin{enumerate}
\item $\ket{\uparrow}\equiv\ket{D_{5/2}, m_j={-3/2}} \rightarrow \ket{S_{1/2}, m_j={+1/2}}$
\item $\ket{S_{1/2}, m_j={+1/2}} \rightarrow \ket{D_{5/2}, m_j={+3/2}}\equiv\ket{\uparrow'}$
\item $\ket{\downarrow}\equiv\ket{S_{1/2}, m_j={-1/2}} \rightarrow \ket{D_{5/2}, m_j={-3/2}}$
\item $\ket{D_{5/2}, m_j={-3/2}} \rightarrow \ket{S_{1/2}, m_j={+1/2}}\equiv\ket{\downarrow'}$
\end{enumerate}

\begin{figure}[!h]
\includegraphics[width=1\columnwidth]{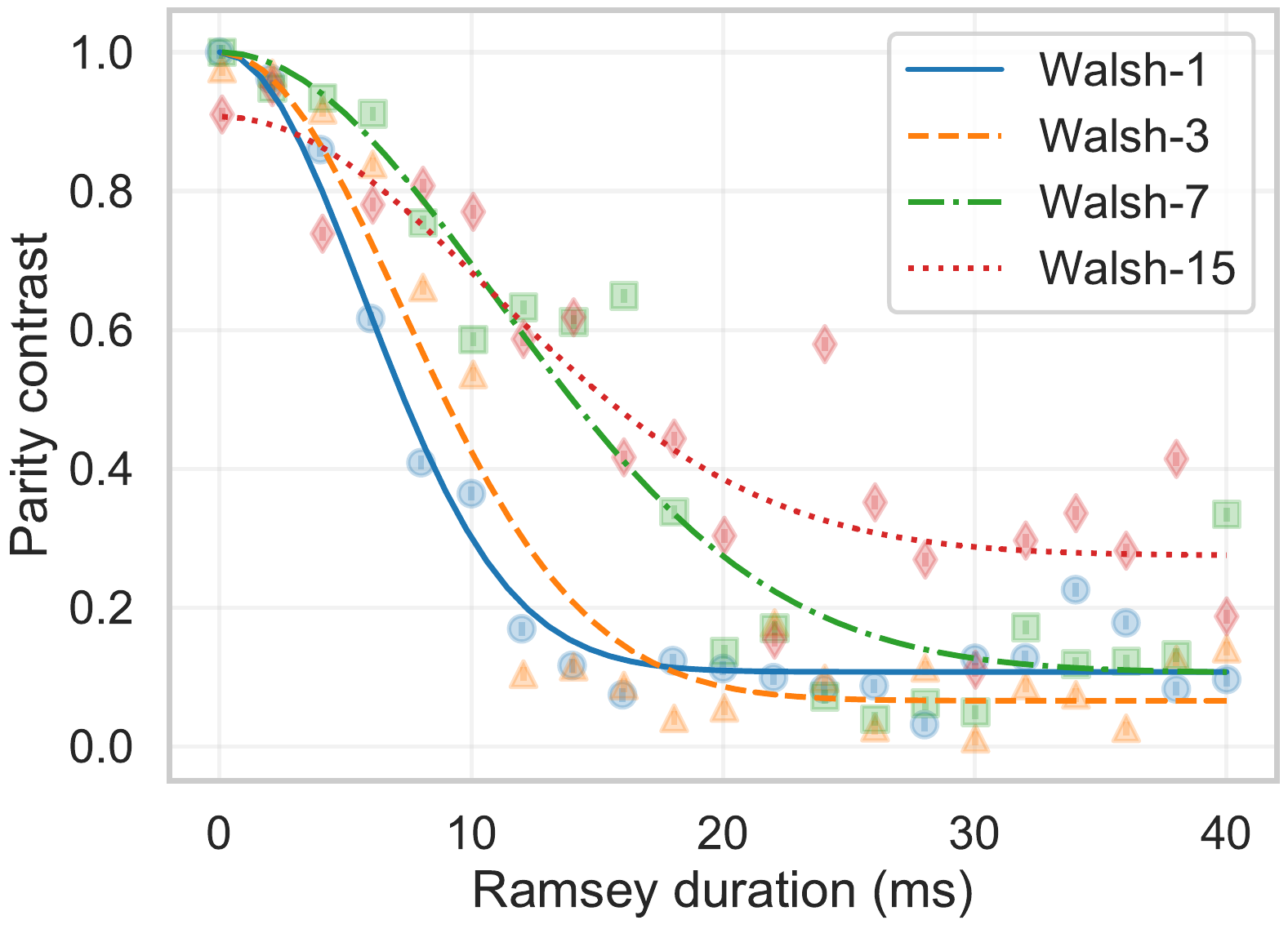}
\centering
\caption{\label{fig_walshes} Single-ion parity contrast versus Ramsey duration for different Walsh sequence indices 1 (blue circles), 3 (orange triangles), 7 (green squares), and 15 (red diamonds). As we increase the Walsh sequence index, we observe higher contrasts at longer durations. We fit a Gaussian decay to obtain the best fit lines. For the experiments in the main text, we use a Walsh-7 sequence.}
\end{figure}

\noindent The total duration of this pulse sequence is $\approx44\,\mu$s. To map $\ket{\downarrow'}\rightarrow\ket{\downarrow}$ and $\ket{\uparrow'}\rightarrow\ket{\uparrow}$, we perform the same sequence in reverse order. For the experiments described in the main text, we perform a total of 5 such sequences to achieve Walsh-7 modulation~\cite{Harmuth1969}. The Ramsey duration $T_R$ excludes the length of these echo sequences. The choice of Walsh sequence index depends on the magnetic field noise spectrum; for our system, the Walsh-7 sequence optimally increased the coherence time without reducing the contrast due to imperfect $\pi$-pulses, as shown in Fig.~\ref{fig_walshes}.

\subsection{Magnetic field stabilisation}
\label{sec_Bstab}

The two primary sources of magnetic-field noise are fluctuations in the current supplied to the coils, and ambient AC fields at 50 Hz (and harmonics) induced by wires carrying mains current in the vicinity of the traps. Stabilisation circuits~\cite{Merkel2019} on the magnetic field coils reduce these noise sources and increase the coherence time of the optical $^{88}\text{Sr}^{+}$ qubit from 0.75(1)\,ms to 7.1(1)\,ms.
 
\subsection{Fibre noise cancellation}
\label{FNC}

\begin{figure*}[t]
{
\includegraphics[width=1.8\columnwidth]{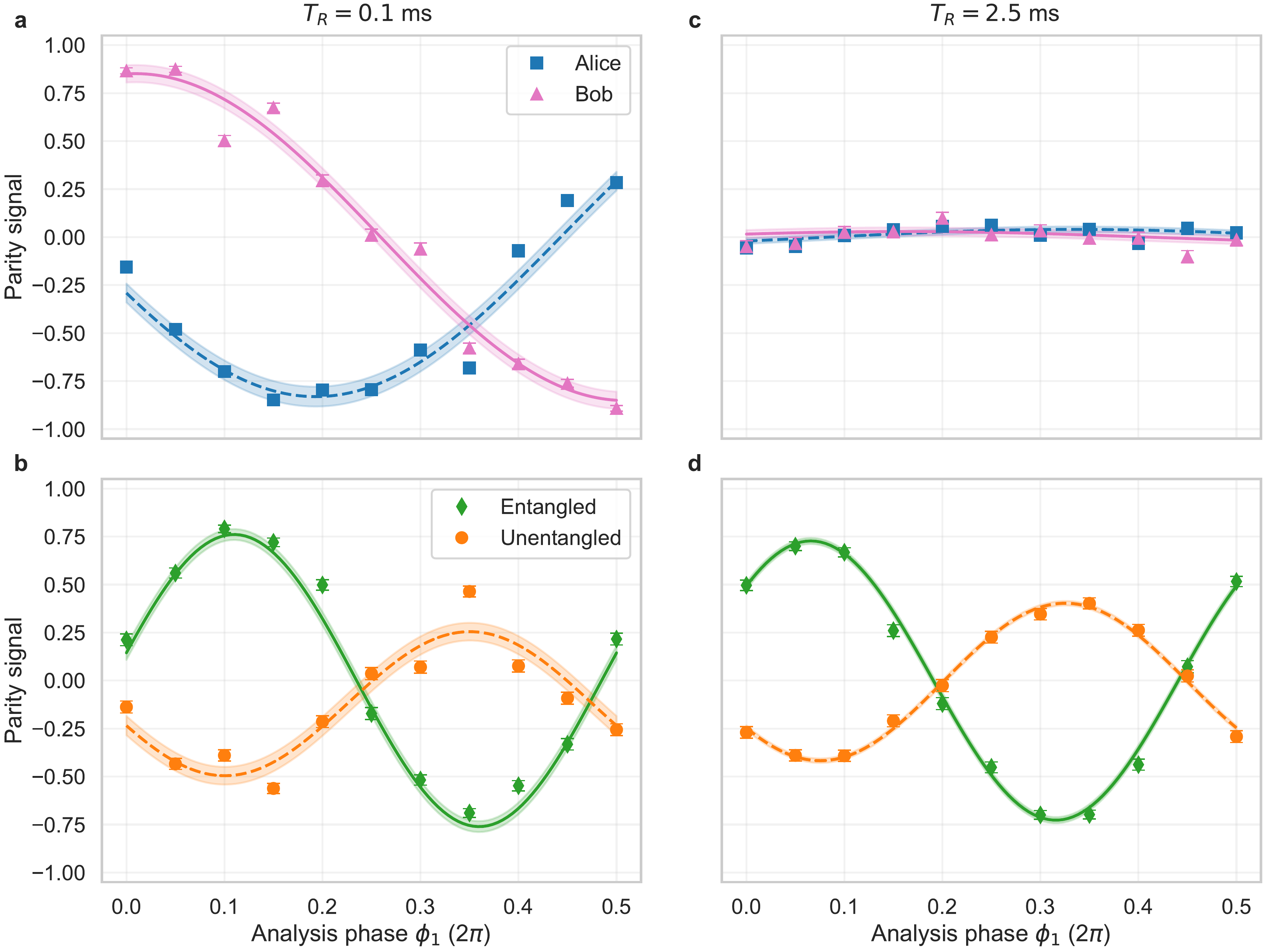}}
\centering
\caption{\label{fig_fnc_off}\textbf{Comparison of parity fringes with fibre noise cancellation (FNC) turned off, with and without entanglement, at 0.1\,ms and 2.5\,ms.} We scan the analysis phase $\phi_1=-\phi_2$ from 0 to $\pi$. We plot the single-ion data for each of Alice (blue squares) and Bob (pink triangles), and the two-ion data with (green diamonds) and without (orange circles) entanglement. \textbf{(a)} and \textbf{(b)} correspond to a Ramsey duration of 0.1\,ms, while \textbf{(c)} and \textbf{(d)} correspond to a Ramsey duration of 2.5\,ms. At 2.5\,ms, qubit decoherence is minimal; the decoherence of the single-ion signals is primarily from laser dephasing, whilst the two-ion signals show no decrease in contrast. The shaded regions show the 68\% confidence intervals of the best fit lines.}
\end{figure*}

The fibre noise cancellation (FNC) for the 674\,nm light uses an acousto-optical modulator (AOM) before the 15\,m fibre to the optical table with Alice and Bob. We pick off light before the AOM, which is superimposed with light retro-reflected from the output facet of the fibre on a photodiode. We use this error signal to modify the rf input to the AOM~\cite{Thirumalai2019}. All the experiments described in the main text were performed with this FNC turned on. The two-ion correlated measurements are insensitive to laser dephasing, as discussed in the text. While this is visible at long Ramsey durations, additional decoherence from magnetic field fluctuations further decreases the parity contrast. To unambiguously demonstrate the insensitivity to laser dephasing, we performed additional experiments with the FNC turned off, as shown in Fig.~\ref{fig_fnc_off}. Compared to the data in Fig.~\ref{fig_parity}, at $T_R=2.5$\,ms we observe a complete loss in contrast of the single-ion signals due to laser dephasing, while the two-ion signals are essentially unchanged.

\subsection{Duration of entanglement generation}
\label{entduration}
\begin{figure}[!h]
{
\includegraphics[width=1\columnwidth]{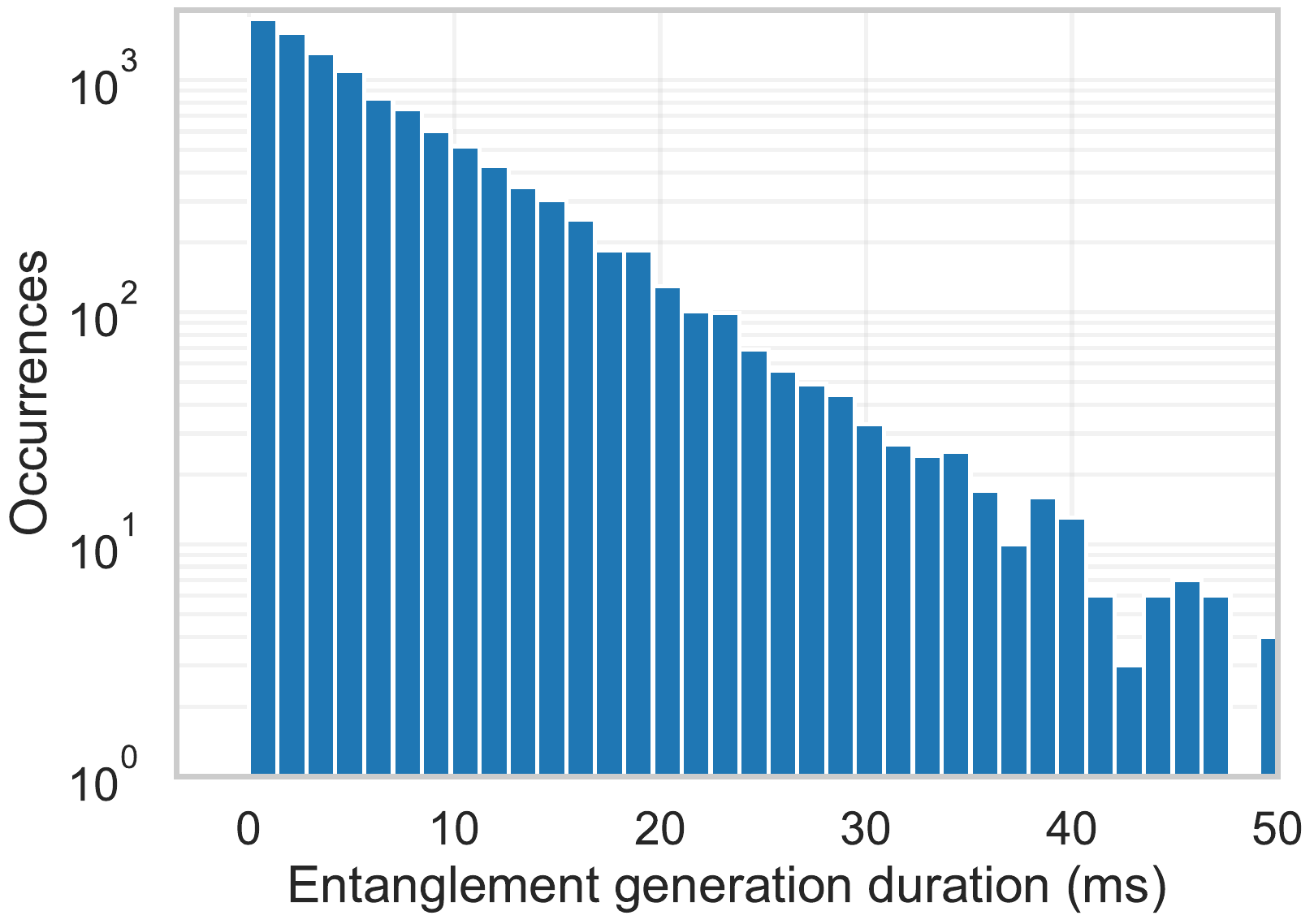}}
\centering
\caption{\label{fig_hist} Histogram of entanglement generation durations for a sample dataset. A total of 11000 attempts are plotted, with a mean duration of 9.2\,ms.}
\end{figure}

We simultaneously excite each ion using a 422\,nm pulse, and the spontaneously emitted photons are transmitted to the Bell state analyser where a coincident detection of a pair of photons projects the ions into an entangled state. As we only detect a small fraction of the emitted photons, the entanglement generation is probabilistic and has a variable duration. We note that, because the entanglement is heralded, all events are used without any post-selection. Each attempt has a duration of 1\,$\mu$s, with 250\,$\mu$s of laser cooling after every 1000 attempts. For the data in this work, the mean entanglement generation duration is about 9\,ms. We plot the histogram of entanglement generation durations for a sample dataset in Fig.~\ref{fig_hist}.

\subsection{Entanglement-enhancement calculation}
\label{enhancement}
As given in Eq.~\ref{eq_single_shot}, the single-shot frequency uncertainty for measuring $\Delta_i=\omega_L-\omega_i$ from a parity scan is

\begin{align*}
    \delta\Delta_i = \frac{\delta\braket{\hat{\Pi}_i}}{C_iT_R},
\end{align*}

\noindent where $T_R$ is the Ramsey wait duration, $C_i$ is the contrast of the single-ion parity fringe, and $\delta\braket{\hat{\Pi_i}}$ is the parity measurement uncertainty, which is ideally limited only by projection noise~\cite{Itano1993} and hence has a value of 1. From the single-ion experiments, the single-shot uncertainty is therefore

\begin{align}
    \delta\Delta_{-,s} = \frac{\delta\braket{\hat{\Pi}_i}}{T_R}\sqrt{\frac{1}{C_1^2}+\frac{1}{C_2^2}}.
\end{align}

\noindent For the correlated two-ion measurements, the single-shot uncertainty is given by

\begin{align*}
    \delta\Delta_{-,u/e} = \frac{\delta\braket{\hat{\Pi}}}{CT_R},
\end{align*}

\noindent where $C$ is now the contrast of the two-ion parity fringe. For $C_1=C_2=C$, using an entangled state gives a $\sqrt2$ reduction in the measurement uncertainty compared to single-atom measurements. Using simultaneous measurements of an unentangled two-ion state, $C$ has a maximum value of $\frac{1}{2}$, which increases the single-shot uncertainty by a factor of 2 compared to the entangled state.

For Fig.~\ref{fig_enhancement}, to obtain an error bar for the single-shot uncertainties and the entanglement enhancement, we only consider the uncertainty in the fitted parity contrasts and propagate the error accordingly.

\subsection{Calibration for Stark shift measurement}
\label{starkcal}
\begin{figure}[H]
{
\includegraphics[width=0.99\columnwidth]{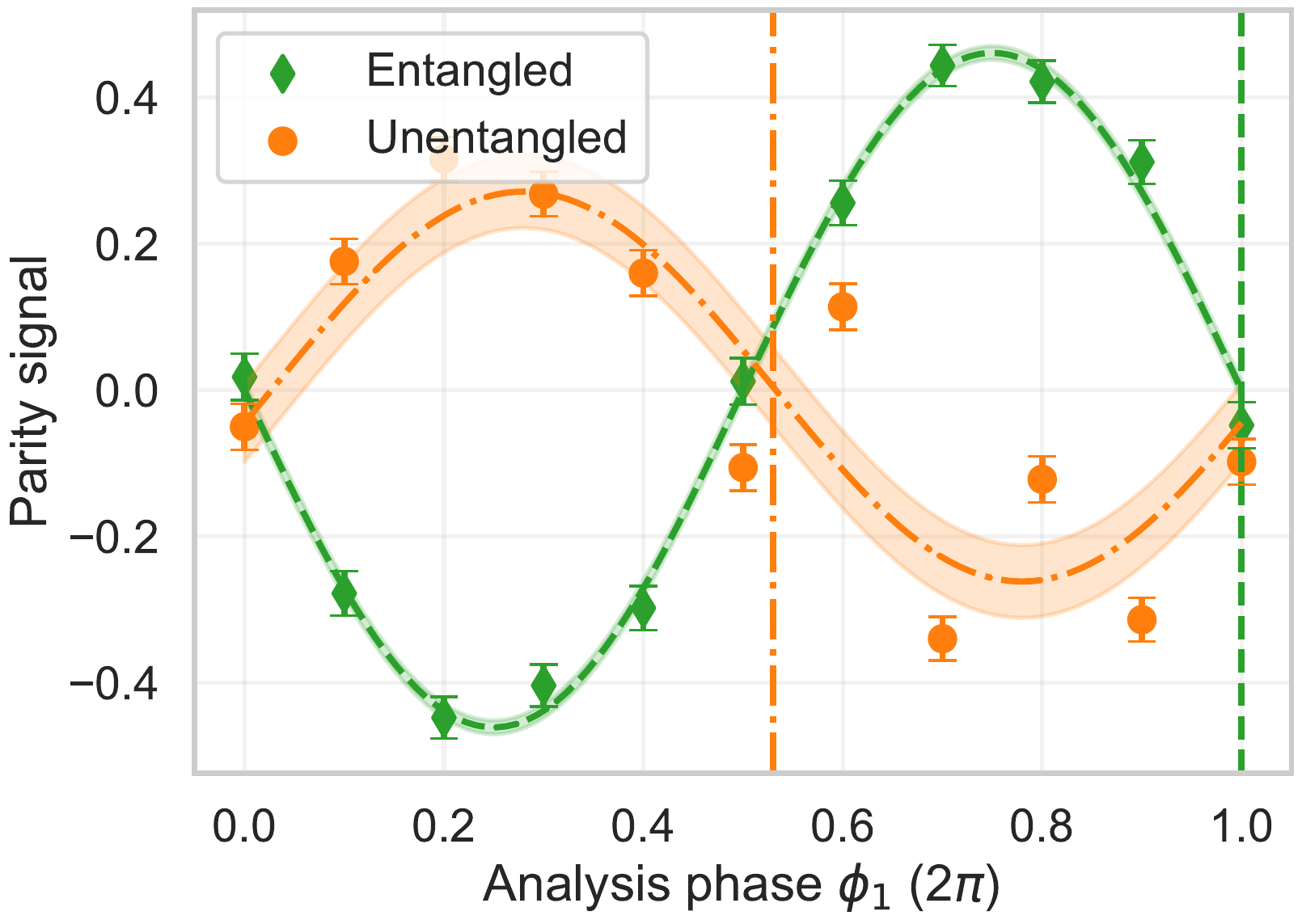}}
\centering
\caption{\label{fig_stark_cal} Calibration data for frequency difference measurement in Fig.~\ref{fig_shift}a. For these data, $\phi_2=0$ and we only scan the value of $\phi_1$ from 0 to $2\pi$ with no shift applied. From these data, we choose the phases at the steepest slopes indicated by the vertical lines for the unentangled (orange dash-dotted) and entangled (green dashed) states, respectively. The shaded regions indicate the 68\% confidence intervals. As in Fig.~\ref{fig_shift}a, the above-statistical scatter of the unentangled data points is evident.}
\end{figure}

To generate a differential shift between the two ions, we illuminate Alice's ion with a detuned 674\,nm beam during the 15\,ms Ramsey wait duration to generate an AC Stark shift. This beam is 35\,MHz red-detuned from the $\ket{S_{1/2}, m_j={-1/2}}\leftrightarrow\ket{D_{5/2}, m_j={-3/2}}$ transition and has a power of $\sim20$\,$\mu$W. The calibration data for the frequency shift measurement in Fig.~\ref{fig_shift} is shown in Fig.~\ref{fig_stark_cal}. The fitted Ramsey fringes allow us to convert a change in the parity signal to a frequency shift. \\
{ }\\

\bibliographystyle{naturemag}
\bibliography{refs}

\begin{thebibliography}{10}
\expandafter\ifx\csname url\endcsname\relax
  \def\url#1{\texttt{#1}}\fi
\expandafter\ifx\csname urlprefix\endcsname\relax\def\urlprefix{URL }\fi
\providecommand{\bibinfo}[2]{#2}
\providecommand{\eprint}[2][]{\url{#2}}

\bibitem{Brewer2019}
\bibinfo{author}{Brewer, S.~M.} \emph{et~al.}
\newblock \bibinfo{title}{{$^{27}{\mathrm{Al}}^{+}$ Quantum-Logic Clock with a
  Systematic Uncertainty below ${10}^{\ensuremath{-}18}$}}.
\newblock \emph{\bibinfo{journal}{Phys. Rev. Lett.}}
  \textbf{\bibinfo{volume}{123}}, \bibinfo{pages}{033201}
  (\bibinfo{year}{2019}).

\bibitem{Oelker2019}
\bibinfo{author}{Oelker, E.} \emph{et~al.}
\newblock \bibinfo{title}{{Demonstration of 4.8$\times$ 10$^{-17}$ stability at
  1 s for two independent optical clocks}}.
\newblock \emph{\bibinfo{journal}{Nature Photonics}}
  \textbf{\bibinfo{volume}{13}}, \bibinfo{pages}{714--719}
  (\bibinfo{year}{2019}).

\bibitem{Ludlow2015}
\bibinfo{author}{Ludlow, A.~D.}, \bibinfo{author}{Boyd, M.~M.},
  \bibinfo{author}{Ye, J.}, \bibinfo{author}{Peik, E.} \&
  \bibinfo{author}{Schmidt, P.~O.}
\newblock \bibinfo{title}{Optical atomic clocks}.
\newblock \emph{\bibinfo{journal}{Rev. Mod. Phys.}}
  \textbf{\bibinfo{volume}{87}}, \bibinfo{pages}{637} (\bibinfo{year}{2015}).

\bibitem{Rosenband2008}
\bibinfo{author}{Rosenband, T.} \emph{et~al.}
\newblock \bibinfo{title}{{Frequency ratio of Al$^+$ and Hg$^+$ Single-Ion
  Optical Clocks; Metrology at the 17th Decimal Place}}.
\newblock \emph{\bibinfo{journal}{Science}} \textbf{\bibinfo{volume}{319}},
  \bibinfo{pages}{1808--1812} (\bibinfo{year}{2008}).

\bibitem{Derevianko2014}
\bibinfo{author}{Derevianko, A.} \& \bibinfo{author}{Pospelov, M.}
\newblock \bibinfo{title}{Hunting for topological dark matter with atomic
  clocks}.
\newblock \emph{\bibinfo{journal}{Nat. Phys.}} \textbf{\bibinfo{volume}{10}},
  \bibinfo{pages}{933--936} (\bibinfo{year}{2014}).

\bibitem{Safranova2018}
\bibinfo{author}{Safronova, M.~S.} \emph{et~al.}
\newblock \bibinfo{title}{Search for new physics with atoms and molecules}.
\newblock \emph{\bibinfo{journal}{Rev. Mod. Phys.}}
  \textbf{\bibinfo{volume}{90}}, \bibinfo{pages}{025008}
  (\bibinfo{year}{2018}).

\bibitem{Mehlstaubler2018}
\bibinfo{author}{Mehlst{\"a}ubler, T.~E.}, \bibinfo{author}{Grosche, G.},
  \bibinfo{author}{Lisdat, C.}, \bibinfo{author}{Schmidt, P.~O.} \&
  \bibinfo{author}{Denker, H.}
\newblock \bibinfo{title}{Atomic clocks for geodesy}.
\newblock \emph{\bibinfo{journal}{Rep. Prog. Phys.}}
  \textbf{\bibinfo{volume}{81}}, \bibinfo{pages}{064401}
  (\bibinfo{year}{2018}).

\bibitem{McGrew2018}
\bibinfo{author}{McGrew, W.} \emph{et~al.}
\newblock \bibinfo{title}{Atomic clock performance enabling geodesy below the
  centimetre level}.
\newblock \emph{\bibinfo{journal}{Nature}} \textbf{\bibinfo{volume}{564}},
  \bibinfo{pages}{87--90} (\bibinfo{year}{2018}).

\bibitem{Meyer2001}
\bibinfo{author}{Meyer, V.} \emph{et~al.}
\newblock \bibinfo{title}{{Experimental Demonstration of Entanglement-Enhanced
  Rotation Angle Estimation Using Trapped Ions}}.
\newblock \emph{\bibinfo{journal}{Phys. Rev. Lett.}}
  \textbf{\bibinfo{volume}{86}}, \bibinfo{pages}{5870--5873}
  (\bibinfo{year}{2001}).

\bibitem{Leibfried2004}
\bibinfo{author}{Leibfried, D.} \emph{et~al.}
\newblock \bibinfo{title}{{Toward Heisenberg-Limited Spectroscopy with
  Multiparticle Entangled States}}.
\newblock \emph{\bibinfo{journal}{Science}} \textbf{\bibinfo{volume}{304}},
  \bibinfo{pages}{1476--1478} (\bibinfo{year}{2004}).

\bibitem{Roos2006}
\bibinfo{author}{Roos, C.~F.}, \bibinfo{author}{Chwalla, M.},
  \bibinfo{author}{Kim, K.}, \bibinfo{author}{Riebe, M.} \&
  \bibinfo{author}{Blatt, R.}
\newblock \bibinfo{title}{{‘Designer atoms’ for quantum metrology}}.
\newblock \emph{\bibinfo{journal}{Nature}} \textbf{\bibinfo{volume}{443}},
  \bibinfo{pages}{316--319} (\bibinfo{year}{2006}).

\bibitem{Megidish2019}
\bibinfo{author}{Megidish, E.}, \bibinfo{author}{Broz, J.},
  \bibinfo{author}{Greene, N.} \& \bibinfo{author}{H\"affner, H.}
\newblock \bibinfo{title}{{Improved Test of Local Lorentz Invariance from a
  Deterministic Preparation of Entangled States}}.
\newblock \emph{\bibinfo{journal}{Phys. Rev. Lett.}}
  \textbf{\bibinfo{volume}{122}}, \bibinfo{pages}{123605}
  (\bibinfo{year}{2019}).

\bibitem{Manovitz2019}
\bibinfo{author}{Manovitz, T.}, \bibinfo{author}{Shaniv, R.},
  \bibinfo{author}{Shapira, Y.}, \bibinfo{author}{Ozeri, R.} \&
  \bibinfo{author}{Akerman, N.}
\newblock \bibinfo{title}{{Precision Measurement of Atomic Isotope Shifts Using
  a Two-Isotope Entangled State}}.
\newblock \emph{\bibinfo{journal}{Phys. Rev. Lett.}}
  \textbf{\bibinfo{volume}{123}}, \bibinfo{pages}{203001}
  (\bibinfo{year}{2019}).

\bibitem{Pedrozo2020}
\bibinfo{author}{Pedrozo-Pe{\~n}afiel, E.} \emph{et~al.}
\newblock \bibinfo{title}{Entanglement on an optical atomic-clock transition}.
\newblock \emph{\bibinfo{journal}{Nature}} \textbf{\bibinfo{volume}{588}},
  \bibinfo{pages}{414--418} (\bibinfo{year}{2020}).

\bibitem{Monroe2014}
\bibinfo{author}{Monroe, C.} \emph{et~al.}
\newblock \bibinfo{title}{Large-scale modular quantum-computer architecture
  with atomic memory and photonic interconnects}.
\newblock \emph{\bibinfo{journal}{Phys. Rev. A}} \textbf{\bibinfo{volume}{89}},
  \bibinfo{pages}{022317} (\bibinfo{year}{2014}).

\bibitem{Stephenson2020}
\bibinfo{author}{Stephenson, L.~J.} \emph{et~al.}
\newblock \bibinfo{title}{{High-Rate, High-Fidelity Entanglement of Qubits
  Across an Elementary Quantum Network}}.
\newblock \emph{\bibinfo{journal}{Phys. Rev. Lett.}}
  \textbf{\bibinfo{volume}{124}}, \bibinfo{pages}{110501}
  (\bibinfo{year}{2020}).

\bibitem{Clements2020}
\bibinfo{author}{Clements, E.~R.} \emph{et~al.}
\newblock \bibinfo{title}{{Lifetime-Limited Interrogation of Two Independent
  $^{27}{\mathrm{Al}}^{+}$ Clocks Using Correlation Spectroscopy}}.
\newblock \emph{\bibinfo{journal}{Phys. Rev. Lett.}}
  \textbf{\bibinfo{volume}{125}}, \bibinfo{pages}{243602}
  (\bibinfo{year}{2020}).

\bibitem{Hume2016}
\bibinfo{author}{Hume, D.~B.} \& \bibinfo{author}{Leibrandt, D.~R.}
\newblock \bibinfo{title}{Probing beyond the laser coherence time in optical
  clock comparisons}.
\newblock \emph{\bibinfo{journal}{Phys. Rev. A}} \textbf{\bibinfo{volume}{93}},
  \bibinfo{pages}{032138} (\bibinfo{year}{2016}).

\bibitem{Komar2014}
\bibinfo{author}{Komar, P.} \emph{et~al.}
\newblock \bibinfo{title}{A quantum network of clocks}.
\newblock \emph{\bibinfo{journal}{Nat. Phys.}} \textbf{\bibinfo{volume}{10}},
  \bibinfo{pages}{582--587} (\bibinfo{year}{2014}).

\bibitem{Wineland1992}
\bibinfo{author}{Wineland, D.~J.}, \bibinfo{author}{Bollinger, J.~J.},
  \bibinfo{author}{Itano, W.~M.}, \bibinfo{author}{Moore, F.} \&
  \bibinfo{author}{Heinzen, D.~J.}
\newblock \bibinfo{title}{Spin squeezing and reduced quantum noise in
  spectroscopy}.
\newblock \emph{\bibinfo{journal}{Phys. Rev. A}} \textbf{\bibinfo{volume}{46}},
  \bibinfo{pages}{R6797} (\bibinfo{year}{1992}).

\bibitem{Wineland1994}
\bibinfo{author}{Wineland, D.~J.}, \bibinfo{author}{Bollinger, J.~J.},
  \bibinfo{author}{Itano, W.~M.} \& \bibinfo{author}{Heinzen, D.}
\newblock \bibinfo{title}{Squeezed atomic states and projection noise in
  spectroscopy}.
\newblock \emph{\bibinfo{journal}{Phys. Rev. A}} \textbf{\bibinfo{volume}{50}},
  \bibinfo{pages}{67} (\bibinfo{year}{1994}).

\bibitem{Degen2017}
\bibinfo{author}{Degen, C.~L.}, \bibinfo{author}{Reinhard, F.} \&
  \bibinfo{author}{Cappellaro, P.}
\newblock \bibinfo{title}{Quantum sensing}.
\newblock \emph{\bibinfo{journal}{Rev. Mod. Phys.}}
  \textbf{\bibinfo{volume}{89}}, \bibinfo{pages}{035002}
  (\bibinfo{year}{2017}).

\bibitem{Caves1981}
\bibinfo{author}{Caves, C.~M.}
\newblock \bibinfo{title}{Quantum-mechanical noise in an interferometer}.
\newblock \emph{\bibinfo{journal}{Phys. Rev. D}} \textbf{\bibinfo{volume}{23}},
  \bibinfo{pages}{1693} (\bibinfo{year}{1981}).

\bibitem{LIGO2019}
\bibinfo{author}{Tse, M.} \emph{et~al.}
\newblock \bibinfo{title}{{Quantum-Enhanced Advanced LIGO Detectors in the Era
  of Gravitational-Wave Astronomy}}.
\newblock \emph{\bibinfo{journal}{Phys. Rev. Lett.}}
  \textbf{\bibinfo{volume}{123}}, \bibinfo{pages}{231107}
  (\bibinfo{year}{2019}).

\bibitem{Malnou2019}
\bibinfo{author}{Malnou, M.} \emph{et~al.}
\newblock \bibinfo{title}{{Squeezed Vacuum Used to Accelerate the Search for a
  Weak Classical Signal}}.
\newblock \emph{\bibinfo{journal}{Phys. Rev. X}} \textbf{\bibinfo{volume}{9}},
  \bibinfo{pages}{021023} (\bibinfo{year}{2019}).

\bibitem{Gilmore2021}
\bibinfo{author}{Gilmore, K.~A.} \emph{et~al.}
\newblock \bibinfo{title}{Quantum-enhanced sensing of displacements and
  electric fields with two-dimensional trapped-ion crystals}.
\newblock \emph{\bibinfo{journal}{Science}} \textbf{\bibinfo{volume}{373}},
  \bibinfo{pages}{673--678} (\bibinfo{year}{2021}).

\bibitem{Kimble2008}
\bibinfo{author}{Kimble, H.~J.}
\newblock \bibinfo{title}{The quantum internet}.
\newblock \emph{\bibinfo{journal}{Nature}} \textbf{\bibinfo{volume}{453}},
  \bibinfo{pages}{1023--1030} (\bibinfo{year}{2008}).

\bibitem{Gisin2002}
\bibinfo{author}{Gisin, N.}, \bibinfo{author}{Ribordy, G.},
  \bibinfo{author}{Tittel, W.} \& \bibinfo{author}{Zbinden, H.}
\newblock \bibinfo{title}{Quantum cryptography}.
\newblock \emph{\bibinfo{journal}{Rev. Mod. Phys.}}
  \textbf{\bibinfo{volume}{74}}, \bibinfo{pages}{145--195}
  (\bibinfo{year}{2002}).

\bibitem{Monroe2013}
\bibinfo{author}{Monroe, C.} \& \bibinfo{author}{Kim, J.}
\newblock \bibinfo{title}{Scaling the ion trap quantum processor}.
\newblock \emph{\bibinfo{journal}{Science}} \textbf{\bibinfo{volume}{339}},
  \bibinfo{pages}{1164--1169} (\bibinfo{year}{2013}).

\bibitem{Hensen2015}
\bibinfo{author}{Hensen, B.} \emph{et~al.}
\newblock \bibinfo{title}{{Loophole-free Bell inequality violation using
  electron spins separated by 1.3 kilometres}}.
\newblock \emph{\bibinfo{journal}{Nature}} \textbf{\bibinfo{volume}{526}},
  \bibinfo{pages}{682--686} (\bibinfo{year}{2015}).

\bibitem{Ramsey1950}
\bibinfo{author}{Ramsey, N.~F.}
\newblock \bibinfo{title}{{A Molecular Beam Resonance Method with Separated
  Oscillating Fields}}.
\newblock \emph{\bibinfo{journal}{Phys. Rev.}} \textbf{\bibinfo{volume}{78}},
  \bibinfo{pages}{695--699} (\bibinfo{year}{1950}).

\bibitem{Ramsey1957}
\bibinfo{author}{Ramsey, N.~F.}
\newblock \bibinfo{title}{Resonance experiments in successive oscillatory
  fields}.
\newblock \emph{\bibinfo{journal}{Rev. Sci. Instrum.}}
  \textbf{\bibinfo{volume}{28}}, \bibinfo{pages}{57--58}
  (\bibinfo{year}{1957}).

\bibitem{Itano1993}
\bibinfo{author}{Itano, W.~M.} \emph{et~al.}
\newblock \bibinfo{title}{Quantum projection noise: Population fluctuations in
  two-level systems}.
\newblock \emph{\bibinfo{journal}{Phys. Rev. A}} \textbf{\bibinfo{volume}{47}},
  \bibinfo{pages}{3554} (\bibinfo{year}{1993}).

\bibitem{Giovannetti2006}
\bibinfo{author}{Giovannetti, V.}, \bibinfo{author}{Lloyd, S.} \&
  \bibinfo{author}{Maccone, L.}
\newblock \bibinfo{title}{{Quantum Metrology}}.
\newblock \emph{\bibinfo{journal}{Phys. Rev. Lett.}}
  \textbf{\bibinfo{volume}{96}}, \bibinfo{pages}{010401}
  (\bibinfo{year}{2006}).

\bibitem{Riis2004}
\bibinfo{author}{Riis, E.} \& \bibinfo{author}{Sinclair, A.~G.}
\newblock \bibinfo{title}{Optimum measurement strategies for trapped ion
  optical frequency standards}.
\newblock \emph{\bibinfo{journal}{J. Phys. B}} \textbf{\bibinfo{volume}{37}},
  \bibinfo{pages}{4719--4732} (\bibinfo{year}{2004}).

\bibitem{Leroux2017}
\bibinfo{author}{Leroux, I.~D.} \emph{et~al.}
\newblock \bibinfo{title}{On-line estimation of local oscillator noise and
  optimisation of servo parameters in atomic clocks}.
\newblock \emph{\bibinfo{journal}{Metrologia}} \textbf{\bibinfo{volume}{54}},
  \bibinfo{pages}{307--321} (\bibinfo{year}{2017}).

\bibitem{Bize2000}
\bibinfo{author}{Bize, S.} \emph{et~al.}
\newblock \bibinfo{title}{Interrogation oscillator noise rejection in the
  comparison of atomic fountains}.
\newblock \emph{\bibinfo{journal}{IEEE Trans. Ultrason. Ferroelectr. Freq.
  Control}} \textbf{\bibinfo{volume}{47}}, \bibinfo{pages}{1253--1255}
  (\bibinfo{year}{2000}).

\bibitem{Chwalla2007}
\bibinfo{author}{Chwalla, M.} \emph{et~al.}
\newblock \bibinfo{title}{Precision spectroscopy with two correlated atoms}.
\newblock \emph{\bibinfo{journal}{Appl. Phys. B}}
  \textbf{\bibinfo{volume}{89}}, \bibinfo{pages}{483--488}
  (\bibinfo{year}{2007}).

\bibitem{Marti2018}
\bibinfo{author}{Marti, G.~E.} \emph{et~al.}
\newblock \bibinfo{title}{{Imaging Optical Frequencies with 100 $\mu$Hz
  Precision and 1.1 $\mu$m Resolution}}.
\newblock \emph{\bibinfo{journal}{Phys. Rev. Lett.}}
  \textbf{\bibinfo{volume}{120}}, \bibinfo{pages}{103201}
  (\bibinfo{year}{2018}).

\bibitem{Young2020}
\bibinfo{author}{Young, A.~W.} \emph{et~al.}
\newblock \bibinfo{title}{{Half-minute-scale atomic coherence and high relative
  stability in a tweezer clock}}.
\newblock \emph{\bibinfo{journal}{Nature}} \textbf{\bibinfo{volume}{588}},
  \bibinfo{pages}{408--413} (\bibinfo{year}{2020}).

\bibitem{Nadlinger2021}
\bibinfo{author}{Nadlinger, D.~P.} \emph{et~al.}
\newblock \bibinfo{title}{{Device-Independent Quantum Key Distribution}}.
\newblock \emph{\bibinfo{journal}{arXiv:2109.14600}}  (\bibinfo{year}{2021}).

\bibitem{Sahoo2006}
\bibinfo{author}{Sahoo, B.~K.}, \bibinfo{author}{Islam, M.~R.},
  \bibinfo{author}{Das, B.~P.}, \bibinfo{author}{Chaudhuri, R.~K.} \&
  \bibinfo{author}{Mukherjee, D.}
\newblock \bibinfo{title}{{Lifetimes of the metastable $^{2}D_{3/2,5/2}$ states
  in ${\mathrm{Ca}}^{+}$, ${\mathrm{Sr}}^{+}$, and ${\mathrm{Ba}}^{+}$}}.
\newblock \emph{\bibinfo{journal}{Phys. Rev. A}} \textbf{\bibinfo{volume}{74}},
  \bibinfo{pages}{062504} (\bibinfo{year}{2006}).

\bibitem{Gabrielse1988}
\bibinfo{author}{Gabrielse, G.} \& \bibinfo{author}{Tan, J.}
\newblock \bibinfo{title}{Self-shielding superconducting solenoid systems}.
\newblock \emph{\bibinfo{journal}{Journal of Applied Physics}}
  \textbf{\bibinfo{volume}{63}}, \bibinfo{pages}{5143--5148}
  (\bibinfo{year}{1988}).

\bibitem{Ruster2016}
\bibinfo{author}{Ruster, T.} \emph{et~al.}
\newblock \bibinfo{title}{{A long-lived Zeeman trapped-ion qubit}}.
\newblock \emph{\bibinfo{journal}{Appl. Phys. B}}
  \textbf{\bibinfo{volume}{122}} (\bibinfo{year}{2016}).

\bibitem{Aharon2019}
\bibinfo{author}{Aharon, N.}, \bibinfo{author}{Spethmann, N.},
  \bibinfo{author}{Leroux, I.~D.}, \bibinfo{author}{Schmidt, P.~O.} \&
  \bibinfo{author}{Retzker, A.}
\newblock \bibinfo{title}{Robust optical clock transitions in trapped ions
  using dynamical decoupling}.
\newblock \emph{\bibinfo{journal}{New J. Phys.}} \textbf{\bibinfo{volume}{21}},
  \bibinfo{pages}{083040} (\bibinfo{year}{2019}).

\bibitem{Schmidt2005}
\bibinfo{author}{Schmidt, P.~O.} \emph{et~al.}
\newblock \bibinfo{title}{{Spectroscopy using quantum logic.}}
\newblock \emph{\bibinfo{journal}{Science}} \textbf{\bibinfo{volume}{309}},
  \bibinfo{pages}{749--752} (\bibinfo{year}{2005}).

\bibitem{Hughes2020}
\bibinfo{author}{Hughes, A.~C.} \emph{et~al.}
\newblock \bibinfo{title}{{Benchmarking a High-Fidelity Mixed-Species
  Entangling Gate}}.
\newblock \emph{\bibinfo{journal}{Phys. Rev. Lett.}}
  \textbf{\bibinfo{volume}{125}}, \bibinfo{pages}{080504}
  (\bibinfo{year}{2020}).

\bibitem{Wright2018}
\bibinfo{author}{Wright, T.~A.} \emph{et~al.}
\newblock \bibinfo{title}{{Two-way photonic interface for linking the Sr$^+$
  transition at 422 nm to the telecommunication C band}}.
\newblock \emph{\bibinfo{journal}{Phys. Rev. App.}}
  \textbf{\bibinfo{volume}{10}}, \bibinfo{pages}{044012}
  (\bibinfo{year}{2018}).

\bibitem{Harmuth1969}
\bibinfo{author}{Harmuth, H.~F.}
\newblock \bibinfo{title}{{Applications of Walsh functions in communications}}.
\newblock \emph{\bibinfo{journal}{IEEE spectrum}} \textbf{\bibinfo{volume}{6}},
  \bibinfo{pages}{82--91} (\bibinfo{year}{1969}).

\bibitem{Merkel2019}
\bibinfo{author}{Merkel, B.} \emph{et~al.}
\newblock \bibinfo{title}{Magnetic field stabilization system for atomic
  physics experiments}.
\newblock \emph{\bibinfo{journal}{Rev. Sci. Instrum.}}
  \textbf{\bibinfo{volume}{90}}, \bibinfo{pages}{044702}
  (\bibinfo{year}{2019}).

\bibitem{Thirumalai2019}
\bibinfo{author}{Thirumalai, K.}
\newblock \emph{\bibinfo{title}{High-fidelity mixed species entanglement of
  trapped ions}}.
\newblock Ph.D. thesis, \bibinfo{school}{University of Oxford}
  (\bibinfo{year}{2019}).

\end{thebibliography}

\end{document}